\documentclass[prd, superscriptaddress, twocolumn,nofootinbib]{revtex4-1}

\usepackage{amsmath}
\usepackage{amssymb}
\usepackage{setspace}
\usepackage{graphicx}
\usepackage{natbib}
\usepackage{subfig}
\usepackage{breqn}
\usepackage{booktabs}
\usepackage{epstopdf}
\usepackage{graphicx}
\usepackage{widetable}

\begin{document}

\title{A proposal for the experimental detection of CSL induced random walk}

\author{\large \bf Sayantani Bera}
\email{sayantani.bera@tifr.res.in} \affiliation{Tata Institute of Fundamental Research\  Homi Bhabha Road\ Mumbai 400005\ India}

\author{\large \bf Bhawna Motwani}
\email{poojauph@iitr.ernet.in} \affiliation{Department of Physics\ Indian Institute of Technology\ Roorkee 247667\ India}

\author{\large \bf Tejinder P. Singh}
\email{tpsingh@tifr.res.in}
\affiliation{Tata Institute of Fundamental Research\ Homi Bhabha Road\  Mumbai 400005\ India}

\author{\large \bf Hendrik Ulbricht}
\email{h.ulbricht@soton.ac.uk}
\affiliation{School of Physics and Astronomy\ University of Southampton\ SO17 1BJ\ UK}

\date{\today}
  
\begin{abstract}
\setstretch{1.1}
\noindent  Continuous Spontaneous Localization (CSL) is one possible explanation for dynamically induced collapse of the wave-function during a quantum measurement. The collapse is mediated by a stochastic non-linear modification of the Schr\"{o}dinger equation. A consequence of the CSL mechanism is an extremely tiny violation of energy-momentum conservation, which can, in principle, be detected in the laboratory via the random diffusion of a particle induced by the stochastic collapse mechanism. In a paper in 2003, Collett and Pearle  investigated the translational CSL diffusion of a sphere, and the rotational CSL diffusion of a disc, and showed that this effect dominates over the ambient environmental noise at low temperatures and extremely low pressures (about ten-thousandth of a pico-Torr). In the present paper, we revisit their analysis and argue that this stringent condition on pressure can be relaxed, and that the CSL effect can be seen at the pressure of about a pico-Torr. A similar analysis is
provided for diffusion produced by gravity-induced decoherence, where the effect is typically much weaker than CSL. We also discuss the CSL induced random displacement of a quantum oscillator.
Lastly, we propose possible experimental set-ups justifying that CSL diffusion is indeed measurable with the current technology.

\end{abstract}

\maketitle
 
\bigskip
\setstretch{1.1}
\section{Introduction}
\noindent The Schr\"{o}dinger equation does not explain the apparent collapse of the wave-function during a quantum measurement, nor the observed absence of macroscopic position superpositions. Since the inception of quantum theory, various explanations have been put forth to explain these observations. These explanations can be broadly divided into two classes. The first class consists of those which modify the interpretation and/or mathematical formulation of quantum theory without altering any of its experimental predictions. These include the Copenhagen Interpretation, Bohmian Mechanics, the Many-worlds Interpretation, Decoherence-based explanations (typically accompanied by additional assumptions such as the environment being an open system, or the many-worlds assumption), and the Consistent Histories formalism. The second class of explanations demonstrate the collapse of the wave-function as a dynamical process by suitable modification of the Schr\"odinger equation as a system approaches the macroscopic regime, while ensuring that the modified equation reproduces all the successful experimental predictions of quantum theory. Prominent amongst this second class are Stochastic non-linear and non-relativistic modifications of the Schr\"odinger equation, such as gravity-induced wave-function collapse, and the model of Continuous Spontaneous Localization (CSL) \cite{PEARLE2, Ghirardi2:90}. For a review, see \cite{Bassi:03,RMP:2012}. The subject of the present paper is a feasibility study for carrying out a possible experimental test to confirm/rule out the CSL model.

The most important prediction of the CSL model is the breakdown of quantum linear position superposition in the limit of approach towards the macroscopic regime. Effectively, what this means is that as objects with large and increasing masses are considered, say for instance an object with mass $10^{9}$ amu, the superposition life-time becomes smaller ultimately rendering the superposition of states essentially unobservable. Thus for high-mass objects, a traditional double-slit experiment would not exhibit fringes, but instead show the classical double hump pattern corresponding to a classical probability distribution. The experimental verification, or otherwise, of this breakdown of superposition predicted by CSL is one of the motives for world-wide ongoing experiments in molecular interferometry, and optomechanics \cite{RMP:2012}. More recently, new ideas for testing and putting bounds on CSL have been proposed - these include tests of CSL-induced spectral line broadening \cite{Bahrami2014} and bounds deduced from heating of an atomic Bose-Einstein Condensate \cite{Pearle2014}. The surge of interest in testing CSL in different ways serves as a premise for witnessing stronger bounds on the CSL model in the coming years.

Another important prediction of the CSL model, which is a consequence of its stochastic nature, is a very tiny violation of energy-momentum conservation. In order that the energy violation does not contradict known physics, significant bounds have been placed on the rate constant $\lambda$ of the CSL model, which is one of the two new fundamental constants introduced in the model, the other being a critical length scale $r_{C}$, assumed to be of the order of $10^{-5}$ cm. In their original work - the GRW model - Ghirardi, Rimini and Weber \cite{Ghirardi:86} assumed that $\lambda_{GRW}$ should be $10^{-16}$ sec$^{-1}$. This is approximately the minimum value required in order to explain the dynamical collapse of a wave-function. The CSL model, which is an improvement over the GRW model, takes $\lambda_{CSL}$ to be\footnote[1]{It has been brought to our attention by Pearle \cite{Pearlepvt} that he prefers $\lambda_{GRW}=\lambda_{CSL}=10^{-16}$ sec$^{-1}$, as used in his paper \cite{PEARLE2}. To our understanding, the paper \cite{Ghirardi2:90} works with $\lambda_{CSL}= 10^{-17}$ sec$^{-1}$.} $10^{-16}$ sec$^{-1}$ or $10^{-17}$ sec$^{-1}$. More recently, Adler has argued, based on analysis and interpretation of latent image formation in photography, that the minimum value of $\lambda$ should be as high as about $\lambda_{ADLER} = 10^{-8}$ sec$^{-1}$. Arguments coming from the non-observation of energy violation set an upper bound on $\lambda$ at roughly $10^{-8}$. The strongest direct experimental upper bound coming from laboratory experiments on interferometry is $10^{-5}$. For a detailed recent discussion on these bounds see \cite{RMP:2012}. 

The tiny energy-momentum violation predicted by CSL also implies that the stochastic kicks experienced by an isolated object will induce a random walk. In principle, under completely ideal conditions, this diffusion should be experimentally detectable. In practice though, such an experiment is extremely difficult and challenging to carry out due to the inevitable presence of various other competing sources of random diffusion. Principal amongst these are (i) thermal Brownian motion (recoil due to emission, absorption and scattering of photons) induced by interaction with photons present in the ambient medium, and (ii) Brownian motion induced by collisions with molecules of the gaseous medium in which the object is immersed. It is also important to note that if the CSL effect does not occur ($\lambda = 0$), intrinsic quantum Brownian motion (time evolution of the expectation value of the position operator) could nonetheless dominate over thermal and gas effects, and care must be taken to avoid mistaking it for the CSL effect.

In an important paper in 2003, Collett and Pearle (CP) ~\cite{PEARLE5} argued that for a particle of suitable size and shape at low temperatures and under extremely low pressures, the CSL diffusion dominates over thermal Brownian motion and gaseous diffusion. Quantum Brownian motion, which would occur if $\lambda = 0$, was shown to be sub-dominant to the CSL diffusion. CP presented their analysis of a CSL translation diffusion for a sphere  and the CSL rotational diffusion for a disc, both having physical dimensions of the order of the CSL localisation length $r_C \sim 10^{-5}$ cm. They showed that, assuming the standard parameter value $\lambda_{GRW}$, the sphere CSL diffuses over a distance of the order of its size in about 20 seconds, and a disc undergoes a rotational CSL diffusion of about $2\pi$ radians in approximately 70 seconds. In order to have the CSL effect dominate over thermal diffusion and gaseous diffusion, CP proposed that experiments could be carried out at the liquid Helium temperature $4.2$ K and an extra-ordinarily low pressure of $< 5 \times 10^{-17}$ Torr. Under these conditions, the mean collision time of air molecules with the sphere/disc is shown to be about 80/45 minutes, consequently allowing adequate time for observation of the CSL effect. In that paper, as well as on subsequent occasions, Pearle has emphasized the importance of carrying out such an experiment. Yet, to the best of our knowledge, an experiment of this kind has not yet been initiated/undertaken. In our opinion, one possible reason for this could be the  extremely low pressures suggested - $10^{-17}$ Torr has been achieved once in the laboratory, however, reproducing the same is as an extra-ordinarily difficult task. This acts as a worthwhile reason for experimentalists to hesitate in pursuing these experiments, despite the fact that in significance, such an experiment significantly rivals the highly successful interferometry experiments for testing the CSL models.

The purpose of our present paper is to revisit the analysis of Collett and Pearle, for three reasons. 
Firstly, we incorporate a more general treatment of the thermal Brownian displacement by including recoil due to emission and absorption of photons in addition to recoil due to scattering. As it turns out, the recoil due to emission is typically dominant over absorption and scattering. Secondly, as has already been emphasized by Adler \cite{Adler3:07}, for the higher value of the fundamental parameter $\lambda_{ADLER}$ proposed by him, the extreme requirement on pressures required for detection of CSL effect is eased. We examine this quantitatively, and show that there is a gain in pressure by almost six orders of magnitude, bringing the value of the new required pressure to around $10^{-11}$ Torr, which in principle is more easily achievable in the laboratory. Thirdly, we observe that the requirement of an extreme pressure of $10^{-17}$ Torr comes about by demanding that the time between collisions of the diffusing particle with air molecules be of the order of tens of minutes, whereas the CSL diffusion time is of the order of tens of seconds (even lesser for the disc, by allowing the observed rotational diffusion of the disc to be in the experimentally measurable range of $\sim 10^{-3}$ instead of $2\pi$ radians). It seems to us that this large ratio (few times 10) between gaseous diffusion time and CSL diffusion time is not necessary from the viewpoint of carrying out a conclusive experiment to detect CSL, and a ratio of $\sim10$ or less is adequate for a plausible statistical analysis of the measurements, and lowers the requirement on pressure to a more feasible value of $\sim10^{-12}$ (pico) Torr, as we demonstrate.

Thus in Sections II and III of this paper, we borrow some of the results of CP, and recalculate the requirements on pressure and temperature, in the light of the two motivations presented in the previous paragraph: the higher value of $\lambda$ argued for by Adler, and the lower ratio between gaseous diffusion time and CSL diffusion which we think should be adequate. We show that the requirement on pressure is considerably eased, making it more likely that an experiment could be carried out. We do this both for the translation of the sphere as well as for the rotation of the disc, and conclude that measurement of the CSL rotation of the disc is a promising experiment to initiate. We carry out the analysis under the assumption that we are in what CP call the `impact realm', where we can talk of individual collisions of the diffusing particle with the air molecules during the time interval of observation. 

The random motion of a localized quantum mechanical particle is influenced by three possible sources: (i) thermal radiation at an ambient temperature $T$, giving rise to thermal Brownian motion, (ii) Brownian motion induced by collision with the molecules of the gas surrounding the localized particle, and (iii) CSL diffusion caused by momentum gain during stochastic wave-function collapse. Given the size and shape of the particle, one can write down a mathematical expression for each of these diffusions. In order for CSL diffusion to be detectable, it should dominate over the thermal motion and over ordinary Brownian motion. Furthermore, it can be shown that the stochastic CSL induced wave-function collapse produces localization of the particle over a time interval shorter than the CSL diffusion time under consideration \cite{PEARLE5}.  Also, intrinsic quantum Brownian motion is shown to be sub-dominant when compared to CSL diffusion \cite{PEARLE5}.

The Brownian motion induced by interaction with the ambient thermal radiation occurs due to absorption, emission, and scattering of photons, which in turn depend on the internal and external temperature of the diffusing object. By requiring that the thermal motion of the particle be a certain fraction of the CSL diffusion in a given time, we fix the internal and external temperature of the object. We then fix the pressure by requiring that the time of measurement of the CSL displacement be shorter than the time between two successive collisions of the particle with gas molecules by a certain factor.  

Gravity induced wave-function decoherence \cite{Karolyhazi:66, Karolyhazi:86, Frenkel:90, Diosi:87} is also known to produce random diffusion, though the effect is considerably weaker than CSL. In analogy with our analysis for CSL, we also work out prospects for detection of gravity induced diffusion, while also emphasising that our estimates for the gravity models are only demonstrative. We do not go into issues relating to additional length cut-offs that need to be introduced in gravity-based collapse models \cite{Diosi:89} so as to avoid conflict with observations. The role of such cut-offs is still an open issue under debate (see for instance the recent discussion in \cite{Bassi2014}) and the estimates provided by us could well change upon a more detailed analysis.

In Section IV, we explore a new system, the CSL ``diffusion" of the quantized oscillator built upon the idea that CSL induces a secular increase in the mean energy of the oscillator \cite{AdlerOsc}. This energy increase translates into a displacement of the mean position of the oscillator, which in classical terms means an enhancement of the amplitude of oscillation. We show that under certain assumptions and suitable conditions, the CSL displacement of the quantum oscillator may be measurable.

In Section V, we put forth possible experimental set-ups, and justify that CSL diffusion is measurable with present technology. In particular, we discuss how the required internal and external temperature as well as very low pressures can be achieved in the laboratory.

In summary, we hope that the considerations we present in this paper will encourage experimentalists to seriously revisit the proposal of CP, consider setting up experiments with the rotational disc and re-examine ongoing quantum oscillator experiments from the viewpoint of detecting CSL diffusion.

\section{The case of a sphere: translational diffusion}

Consider a quantum mechanical spherical object of radius $R$, whose wave-packet is assumed to be localized by the CSL mechanism, and which is immersed in an ambient gaseous medium of temperature $T_e$ and pressure $P$.

Collett and Pearle show that the thermal displacement of a sphere of density $D$ (expressed in gms/cc) is given at temperature $T_e$ by equation 4.9 of ~\cite{PEARLE5} (hereafter referred to as CP) as -
\begin{equation}
\Delta x_{RAD,CP} \approx 8D^{-1} (T_e/T_0)^{9/2}(t/10^5)^{3/2} \rm{cm}
\label{sphere_thermal}
\end{equation}
where $t$ is in seconds and $T_0=300K$ is the room temperature.
This effect however has been calculated solely on the basis of recoil induced due to scattering of photons, in effect, assuming that the object is a perfect reflector/transmitter. A more general treatment would take into account also the recoil due to emission and absorption, as discussed for instance in \cite{Romero2012decoherence}. For our analysis, we estimate the thermal displacement of a sphere according to this more general treatment.

Assuming a start from rest at time $t=0$, the thermal displacement at time $t$ is given by \cite{Romero2012decoherence} 
\begin{equation}
\langle \hat{x}^2(t)\rangle = \frac{2\Lambda\hbar^2}{3M^2} t^{3}
\end{equation} 
where $\Lambda$ is the so-called localization parameter which includes contributions from scattering, emission and absorption of thermal photons. These contributions are respectively denoted by
$\Lambda_{sc}$, $\Lambda_e$ and $\Lambda_a$, so that $\Lambda= \Lambda_{sc} + \Lambda_e +\Lambda_a$, with the individual components being given by
\begin{equation}
\Lambda_{sc} = \frac{8!\times 8\;\xi(9) cR^6}{9\pi} \left [ \frac{k_BT_e}{\hbar c}\right]^9{\rm Re}\left[\frac{\epsilon -1}{\epsilon + 2}\right]^2
\end{equation} 
\begin{equation}
\Lambda_{e(a)} = \frac{16\pi^5 cR^3}{189} \left [ \frac{k_BT_{i(e)}}{\hbar c}\right]^6{\rm Im}\left[\frac{\epsilon -1}{\epsilon + 2}\right]^2
\end{equation} 
Here, $\xi$ is the Riemann-zeta function and the dielectric constant $\epsilon$ is assumed to be of the order of unity. In the subsequent estimates, the contribution from the real and imaginary parts of the fraction $(\epsilon-1)/(\epsilon+2)$ on the right hand side of the above two equations will be set to one. The internal temperature $T_i$ is a measure of the internal energy (coming from rotational and vibrational degrees of freedom) of the bulk object and will in general be different from the ambient radiation temperature $T_e$. The internal temperature plays an important role in the following discussion, and taking it into account makes our analysis different from that of Collett and Pearle.

In order to get a fair estimate of the relative importance of emission-induced recoil with respect to scattering and absorption, we take ratios from the above two equations (ignoring numerical coefficients). From this, we obtain
\begin{equation}
\frac{\Lambda_e}{\Lambda_{sc}}\sim\left(\frac{\hbar c/k_B T_e}{R}\right)^3 \left(\frac{T_i}{T_e}\right)^6,\frac{\Lambda_a}{\Lambda_{sc}}\sim\left(\frac{\hbar c/k_B T_e}{R}\right)^3
\end{equation}
The thermal wavelength $\hbar c / k_B T_e$ is of the order 1 cm for a temperature $T_e = 1 K$, and we will work with a particle size $R\sim 10^{-5}$ cm. Thus we can rewrite these ratios as
\begin{equation}
\frac{\Lambda_e}{\Lambda_{sc}} \sim \left(\frac{10^5}{T_e}\right)^3\; \left(\frac{T_i}{T_e}\right)^6,\quad\frac{\Lambda_a}{\Lambda_{sc}} \sim  \left(\frac{10^5}{T_e}\right)^3
\label{lambda_values}
\end{equation}
It is evident that if the external temperature is equal to or less than the room temperature, scattering can be neglected in comparison to absorption. Furthermore, if the internal temperature is greater than (of the order of) the external temperature, emission dominates (is of the order of) absorption. Hence, in the following estimates, we will only consider recoil due to emission, and effectively set $\Lambda=\Lambda_e$.

Thus, the radiation induced displacement may now be written as
\begin{equation}
\Delta x_{RAD}= \sqrt{\frac{2}{3}}\Lambda_e^{1/2}\frac{\hbar}{M} t^{3/2}
\end{equation}
and using the expression for $\Lambda_e$ from equation (\ref{lambda_values}) we get
\begin{equation}
\Delta x_{RAD}= 6.35\times 10^{-20} D^{-1}R^{-3/2}T_i^3 t^{3/2}
\label{rad}
\end{equation}
where $D$ is the density of the particle in gms/cc. We note that this expression differs from the one due to CP (as depicted above in equation(\ref{sphere_thermal})). 

Now if we fix $T_e$ at the room temperature i.e. $300$ K, we find that the lowest permissible value of $T_i$ is $\sim 76.3$ K. If $T_i$ is less than this then the scattering effect becomes stronger than emission. On the other hand, if we take $T_e = 100$ K, then we can lower $T_i$ upto $14.7$ K. In the following calculation, we shall fix the external temperature $T_e$ at $100$ K.

The CSL-induced translational diffusion, according to equation 4.5b of CP, is given by
\begin{equation}
\Delta x_{CSL} = \lambda^{1/2}\left[\frac{\hbar^2f t^3}{6m^2a^2}\right]^{1/2} = 20 \lambda^{1/2} t^{3/2}
\end{equation}
where $t$ is in seconds and $m$ is the mass of a nucleon. The symbol $a$ stands for the CSL critical length, which as we noted above, is usually denoted by $r_C$. The function $f(R/a)$ has an analytic form as in equation A9b of CP, and is equal to 0.62 for $R=a$. Also, here we have set $R=a=10^{-5}$ cm.

\begin{figure}
\includegraphics[width=\columnwidth]{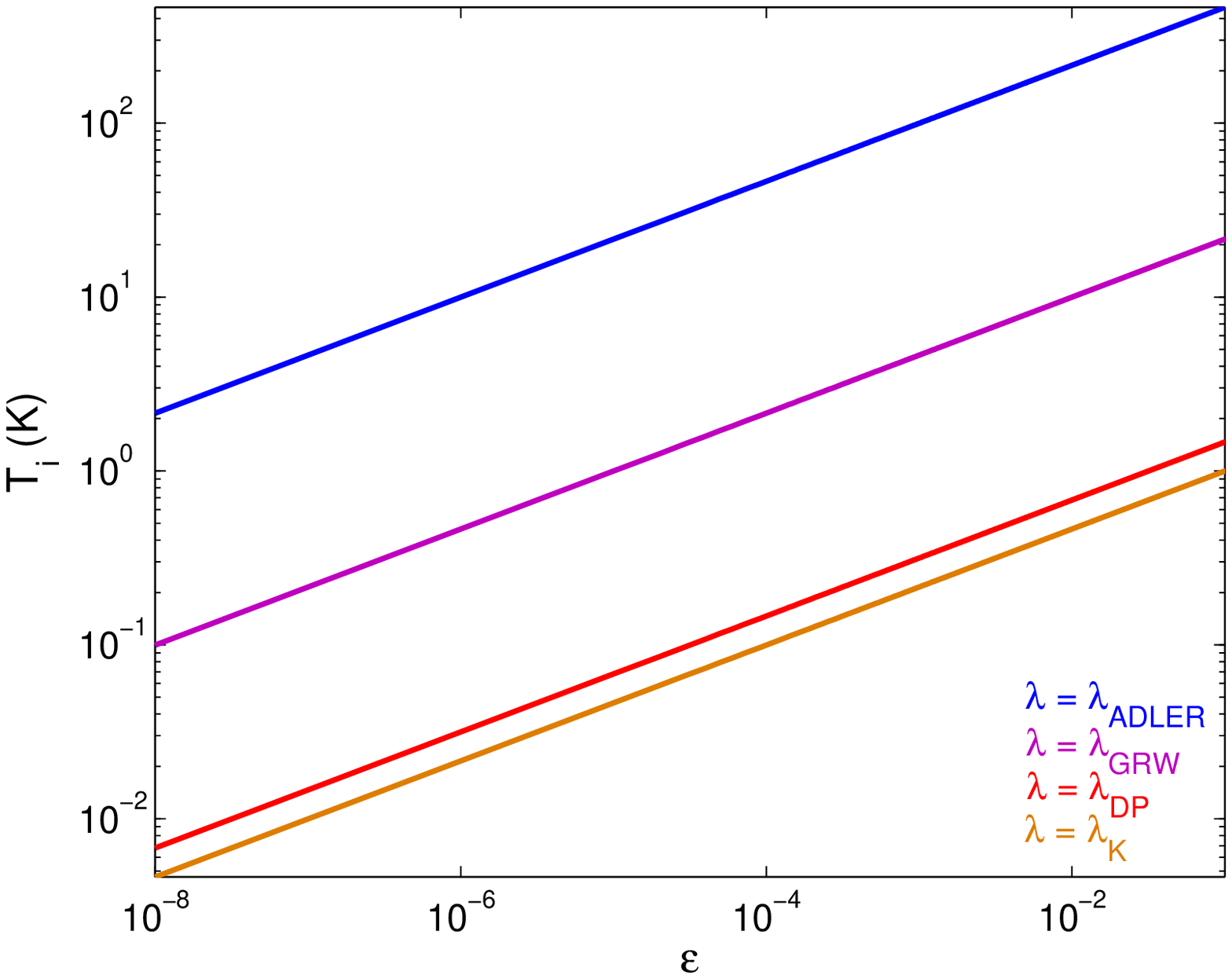}
\includegraphics[width=\columnwidth]{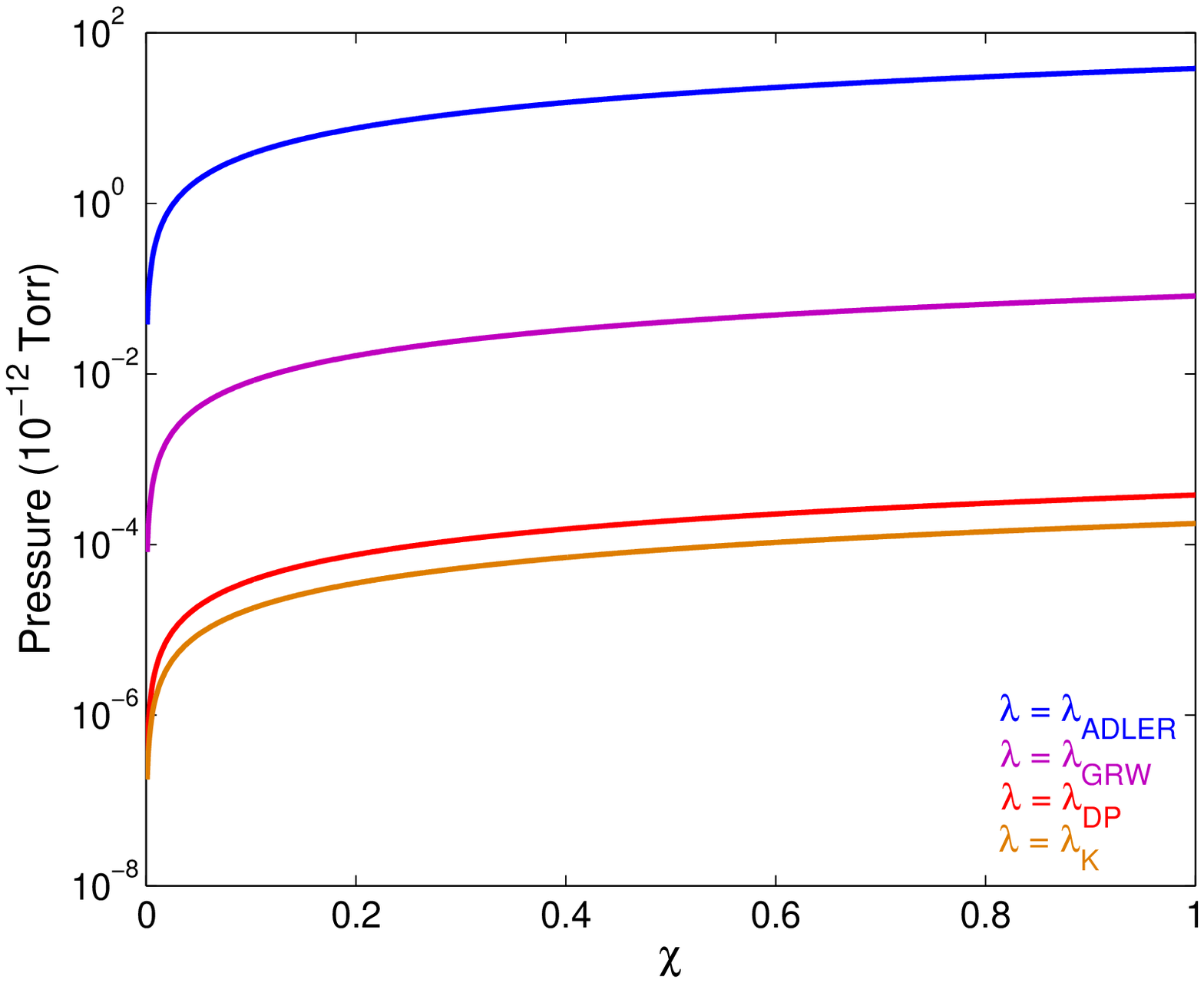}
\includegraphics[width=\columnwidth]{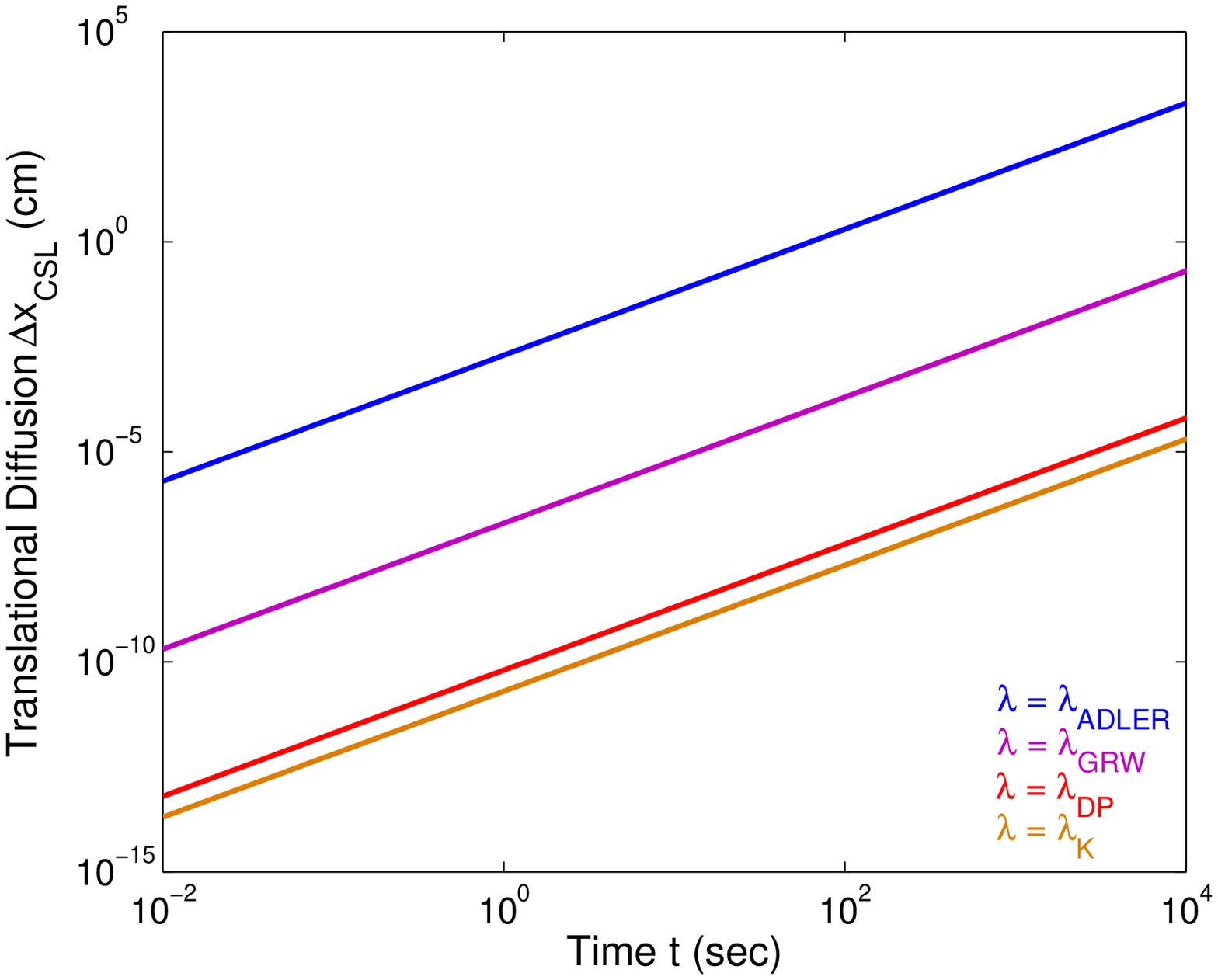}
\caption{\raggedright CSL diffusion of a Sphere: Relative significance of thermal and CSL displacements constrains the  internal temperature $T_i$, plotted here  vs. the fraction $\epsilon = \Delta x_{RAD}/ \Delta x_{CSL}$ [top]; Relative significance of gas collisions and CSL displacement constrains required pressure  $P$ plotted here vs. the fraction $\chi=t_{CSL}/\tau_c$ [middle]; CSL translational displacement $\Delta x_{CSL}$ vs. time for different models [bottom]. The different labelings on the rate constant $\lambda$ denote the different choices: ADLER, GRW,  and the two gravity models by Diosi-Penrose and Karolyhazy}
\label{sphere}
\end{figure}

By demanding that the radiation-induced diffusion $\Delta x_{RAD}$ be a fraction $\epsilon$ of the CSL-induced translational diffusion $\Delta x_{CSL}$, we see that
\begin{equation}
T_i \approx 6.8 \times 10^6 (\epsilon D)^{1/3} R^{1/2} \lambda^{1/6}
\label{cslsphtemp}
\end{equation}
Using $D$=1gm/cc, $R=10^{-5}$ cm and $\epsilon =0.1$, we obtain
\begin{eqnarray}
\nonumber T_i\: &&= 21.5 \: \textrm{K} \quad \quad \quad for \quad \lambda_{GRW} =10^{-16} \;{\rm{sec}}^{-1}\\
&&= 463 \: \textrm{K} \quad \quad \quad for \quad \lambda_{ADLER} =10^{-8} \;{\rm{sec}}^{-1}
\label{Tintsph}
\end{eqnarray}
The top panel of figure \ref{sphere} shows the dependence of internal temperature $T_i$ on the fraction $\epsilon$ for different values of model parameter $\lambda$. 

At an external temperature $T_e$ and pressure $P$, the mean time between two molecule-sphere collisions is given, in the impact realm, by (equation 4.6 of CP)
\begin{equation}
\tau_c \approx 2(T_e/T_0)^{1/2}( P\: pT)^{-1} \quad \text{sec}
\end{equation} 
where the pressure is given in pico Torr ($pT$).

If a CSL displacement $\Delta x_{CSL}\equiv l$ takes place in a time $t_{CSL}$, we want $t_{CSL}$ to be a fraction $\chi$ of $\tau_c$. Therefore,
\begin{equation}
P\: = 0.82\:T_e^{1/2} \chi \lambda^{1/3}\frac{1}{\Delta x_{CSL}^{2/3}}\quad pT
\label{pressure}
\end{equation}
For $T_e =100$ K, $\chi=0.1$ and $\Delta x_{CSL} =10^{-5}$ cm we have,
\begin{eqnarray*}
P\: &&= 8.2\times 10^{-3} \: pT \quad \quad \quad for \quad \lambda_{GRW} =10^{-16} \;{\rm{sec}}^{-1}\\
&&= 3.8 \: pT  \quad \quad \quad \quad \quad \quad for \quad \lambda_{ADLER} =10^{-8} \ {\rm{sec}}^{-1}\\
\end{eqnarray*}

\begin{table*}
\begin{tabular}{l l l l l l l} 
\hline
$\Delta x_{CSL} (cm)$ \quad \quad & $t_{CSL}$ for $\lambda_{GRW}$ \quad \quad & $t_{CSL}$ for $\lambda_{ADLER}$ \quad \quad & $t_{Karolyhazy}$ \quad \quad \quad & $t_{DP}$ \quad \quad \quad \quad &$t_{QBD}$ \\ [1ex]
\hline                  
$10^{-5}$ & 13 & 0.03 & $6.3\times 10^3$ & $ 3\times 10^3$ & $17 \times 10^{2}$  \\ 
$10^{-4}$ & 63 & 0.13 & $3\times 10^4$ & $1.4\times 10^4 $ & $17 \times 10^{3}$ \\ 
$10^{-3}$ & 292 & 0.6 & $1.4\times 10^5$ & $6.3\times 10^4 $ & $17 \times 10^{4}$ \\ 
$10^{-2}$ & 135 & 3 & $6.3\times 10^5 $ & $3\times 10^5 $ & $17 \times 10^{5}$ \\ [1ex]
\hline 
\label{table1}
\end{tabular}
\caption{Displacement time $t_{CSL}$ for a sphere (R = a) in $sec$, for different displacement values and models. The time for quantum Brownian motion $t_{QBD}$ exceeds $t_{CSL}$ and $t_{GRW}$ but is comparable to or dominant over the displacement time in gravity models.}
\end{table*}

The time $t_{CSL}$ for $\lambda_{ADLER}$ and $\Delta x_{CSL} = 10^{-5}$ cm is about $10^{-2}$ sec.
Table I shows the values of $t_{CSL}$ for a few different choices of parameters. Figure \ref{sphere} middle panel shows the dependence of pressure on fraction $\chi$. The quantum Brownian motion of the sphere can be calculated in accordance with equation (3.4) of CP and the discussion in section IV(A) of CP. The same is displayed for select parameter values in Table I and is clearly less important than CSL diffusion.

Thus, in a nutshell, we notice the following significant difference from the inferences of CP: by taking the thermal displacement to be a tenth of the CSL displacement, the CSL displacement itself to be about $10^{-5}$ cm, the CSL diffusion time as one-tenth of the gaseous collision time (rather than something much lower), and $\lambda$ equal to $\lambda_{ADLER}$, we get the required external temperature to be $\sim$100 K, internal temperature $\sim$400 K, and the required pressure to be about $10^{-12}$ Torr. These appear to be feasible choices for an experiment, achievable with current technology.

\subsection*{Gravity Induced Diffusion}
In the model of Karolyhazy, the gravity induced displacement of an isolated solid object, after it has performed a large number of expansion-reduction cycles, is given by \cite{Frenkel:90}
\begin{equation}
\Delta x_{grav}\sim \frac{1}{10}a_c \Big(\frac{t}{\tau_g}\Big)^{3/2}
\label{Kdisp}
\end{equation}
where $a_c$ denotes the critical coherence cell length and $\tau_g$ is the corresponding decoherence time. By demanding this to be a fraction $\epsilon$ of the thermal displacement, we get the internal temperature to be
\begin{equation}
T_i=1.16\times 10^6 (\epsilon D)^{1/3}\sqrt{\frac{R}{\tau_g}}a_c^{1/3}
\end{equation}
Now, for an object of density $D=1$ gm/cc and radius $R=10^{-5}$ cm, we have $a_c=10^{-5}$ cm, $M= 10^{-14}$ gms and $\tau_g \sim 1000 s$ and this is the micro-macro transition region in the Karolyhazy model.
Putting these values in the above equation, we get, for $\epsilon =0.1$ and $D=1$ gm/cc,
$T_i = 1.16 K$. This means that to test the gravity models, we cannot keep $T_e$ as high as $300$ K because then scattering will dominate. So we assume a lower value, say $T_e=1$ K.
 
 As before, by writing $t_{grav}=\chi \tau_c$ where $\tau_c$ is the mean molecule-sphere collision time, we find the pressure to be
 \begin{equation}
P(pT) =0.03\chi T_e^{1/2}\Big(\frac{a_c}{\Delta x_{grav}}\Big)^{2/3} \frac{1}{\tau_g}
\end{equation}
From this expression, again using the same parameter values, we get, for $\Delta x_{grav}= 10^{-5}$ cm, $T_e=1$ K and $\chi =0.1$, 
\begin{equation}
P= 3 \times 10^{-6} pT
\end{equation}
which is an extremely stringent requirement on the pressure.

This result can also be obtained by first working out an effective value $\lambda=\lambda_K$ for the Karolyhazy model, by comparing it to CSL, and then using $\lambda=\lambda_K$ in the above CSL analysis. For this, we note that $\Delta x_{grav}$ has the same time dependence as $\Delta x _{CSL}$, and comparing the two we can write  
 \begin{equation}
20 \lambda_{grav}^{1/2}t^{3/2}=\frac{1}{10}a_c \Big(\frac{t}{\tau_g}\Big)^{3/2}
\end{equation}
Putting $R=a_c=10^{-5}$ cm and $\tau_c =1000$ sec, we get,
\begin{equation}
\lambda_{K}=10^{-24} \; {\rm sec}^{-1}
\end{equation}
This value represents an effective $\lambda$ parameter equivalent to the CSL $\lambda$ parameter for an object of density 1 gm/cc, and is significantly lower than the CSL value.

A similar calculation can be done for the Di\'osi-Penrose model of gravitational decoherence. Collett and Pearle  in their paper \cite{PEARLE5} discuss the case of gravity induced diffusion using Di\'osi-Penrose model (see their appendix E). By comparing the equilibrium size of a wave packet with CSL results, they have calculated an effective $\lambda$ for the Di\'osi-Penrose model as $\lambda_{DP}=Gm^2/a\hbar$, where $m$ is the nucleon mass. Taking $a=10^{-5}$ cm they estimate the effective value to be $\lambda_{DP} \sim 10^{-23} sec^{-1}$.
If we use this value in $\eqref{cslsphtemp}$ and $\eqref{pressure}$ keeping the other parameters same, then the temperature and pressure for the translational diffusion of a sphere come out to be
$T_i \sim 1.5 K $,  $P \sim 3.8 \times 10^{-6} pT$.

Figure 1 bottom panel shows the relative magnitudes of the CSL type random diffusion for different models. Clearly, gravity induced diffusion is considerably weaker than the CSL effect.

We observe from Table I that quantum Brownian displacements are comparable to or dominant over gravity diffusion, suggesting that even at such low pressures it may not be possible to detect gravity induced random walk. It is important to know the magnitude of the quantum Brownian motion which would occur if CSL were to be false. For instance, if the two Brownian motions were comparable (CSL and pure quantum) then a detection would not be able to discriminate between the two. In the case of gravity induced decoherence, the quantum Brownian motion could be suppressed by going to a higher mass, since it scales inversely with mass. Thus if we say raise the size by one order of magnitude, to $10^{-4}$ cm, the mass goes up by three orders, to $10^{-12}$ gms [$10^{12}$ amu]. It can be shown for the Karolyhazy model that $\lambda_{grav}$ changes very weakly; hence gravitational diffusion is not significantly affected. The required pressure goes down by another order of magnitude. Thus the quantum  Brownian motion is sufficiently suppressed - by three orders; and in principle gravity diffusion can be detected by going to very very low pressures (about $10^{-7}$ pT). Similar conclusions hold for the DP model.

\section{The case of a disc: rotational diffusion}
We consider next the rotational diffusion of a suspended disc of radius $R$ and thickness $b\ll L$.
The thermal angular displacement for a disc can be estimated by taking the translational thermal diffusion expression for the sphere from Eqn. (\ref{rad}) and dividing by the radius $L$ of the disc, to get:
\begin{equation}
\Delta\theta_{RAD}=6.35\times 10^{-15} D^{-1}R^{-3/2}T_i^3 t^{3/2}
\end{equation}

The CSL rotational diffusion of the disc is given by (CP equation 6.5)
\begin{equation}
\Delta\theta_{CSL} \approx 0.018 f^{1/2}_{ROT}t^{3/2}\times \frac{\lambda^{1/2}}{10^{-8}} 
\end{equation}
where $f_{ROT}(\gamma,\beta)$ is a function of $\gamma\equiv L/2a$ and $\beta\equiv b/2a$, $b$ being the width of the disc. For $b\approx 0.5a$ and $L\approx 2a$, $f_{ROT}\approx 1/3$, and here we work with this value.
  
By assuming $\Delta\theta_{RAD} = \epsilon \Delta\theta_{CSL}$ we get for the temperature
\begin{equation}
T_i= 5.47\times 10^6 (\epsilon D)^{1/3} R^{1/2} \lambda^{1/6}
\label{disctemp}
\end{equation}
Again for the same parameters and $\epsilon=0.1$ we get,
\begin{eqnarray}
\nonumber T_i\: &&= 17\: \textrm{K} \quad \quad \quad for \quad \lambda_{GRW} =10^{-16} \; {\rm sec}^{-1}\\
&&= 365\: \textrm{K} \quad \quad \quad for \quad \lambda_{ADLER} =10^{-8} \; {\rm sec}^{-1}
\label{Tintdisc}
\end{eqnarray}
Fig. 2 top panel shows the dependence of internal  temperature on the fraction $\epsilon$.

The mean time between two molecule-disc collisions is given, in the impact realm, as above (CP equation 6.7)
\begin{equation}
\tau_c \approx 1.03(T_e/T_0)^{1/2} ( P\ pT)^{-1} \rm{sec}
\end{equation} 
By repeating the calculation as in the case of the sphere, we find the required pressure as
\begin{equation}
P (pT)=616 \chi T_e^{1/2} \lambda^{1/3} \frac{1}{\Delta\theta_{CSL}^{2/3}}
\label{discpressure}
\end{equation}
Keeping other parameters same as before and taking $\Delta\theta_{CSL}= 1$ milli-radian, we get,
\begin{eqnarray*}
P &&= 0.3\: pT \quad \quad  for \quad \lambda_{GRW} =10^{-16} \;{\rm sec}^{-1}\\
&&= 132.7\: pT  \quad \quad  for \quad \lambda_{ADLER} =10^{-8} \;{\rm sec}^{-1}\\
\end{eqnarray*}

In the case of a rotating disc, and for $\lambda=\lambda_{ADLER}$, the required pressure is about 
$100 pT$, which is clearly a much favourable situation compared to the sphere. The time $t_{CSL}$ for $\lambda_{GRW}$ is 0.2 sec, and for $\lambda_{ADLER}$ it is $4 \times 10^{-4}$ sec. Figure 2 middle panel shows the dependence of the pressure on the fractions $\chi$, and Table II shows $t_{CSL}$ for a range of parameters.
 
\begin{figure}
\includegraphics[width=\columnwidth]{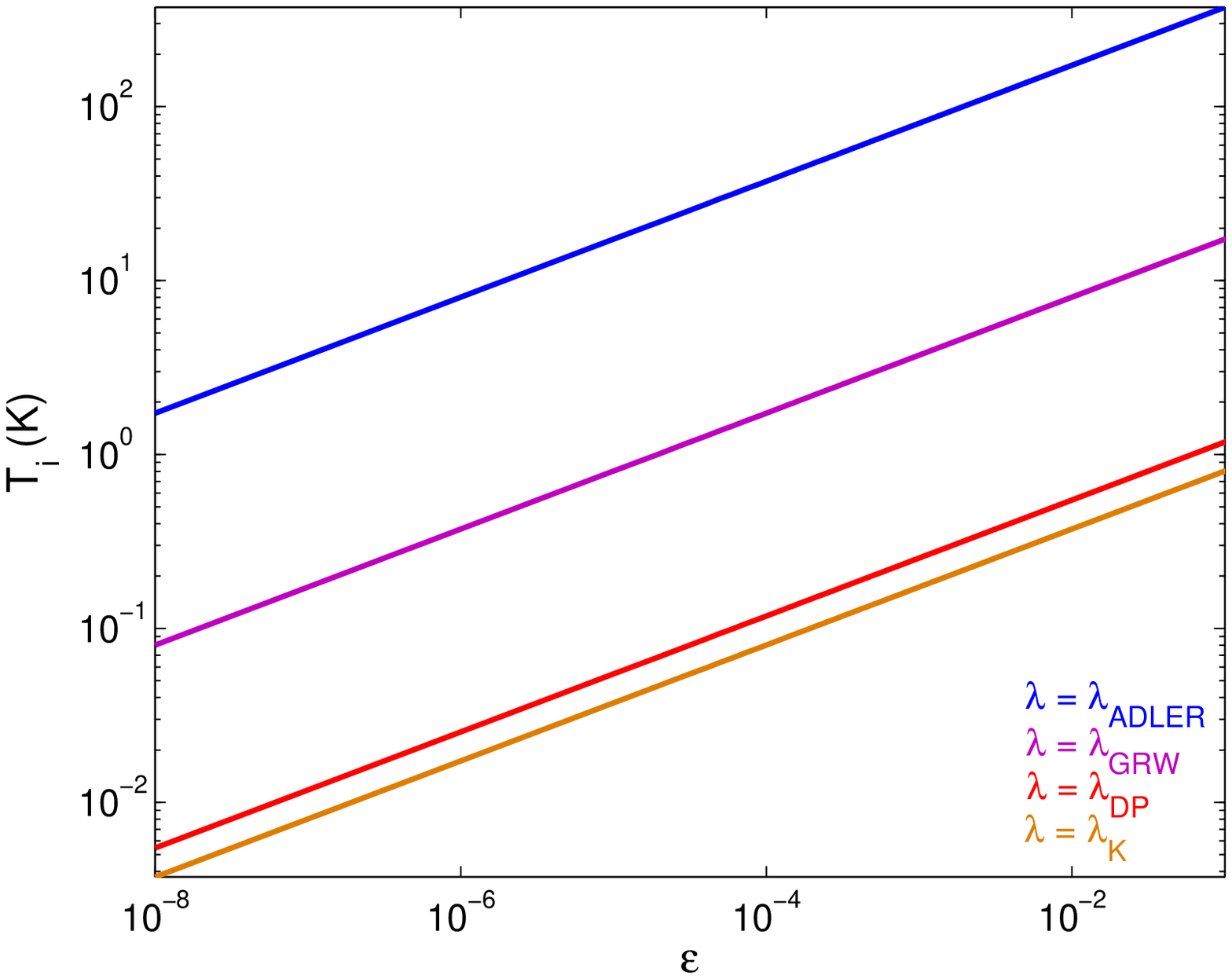}
\includegraphics[width=\columnwidth]{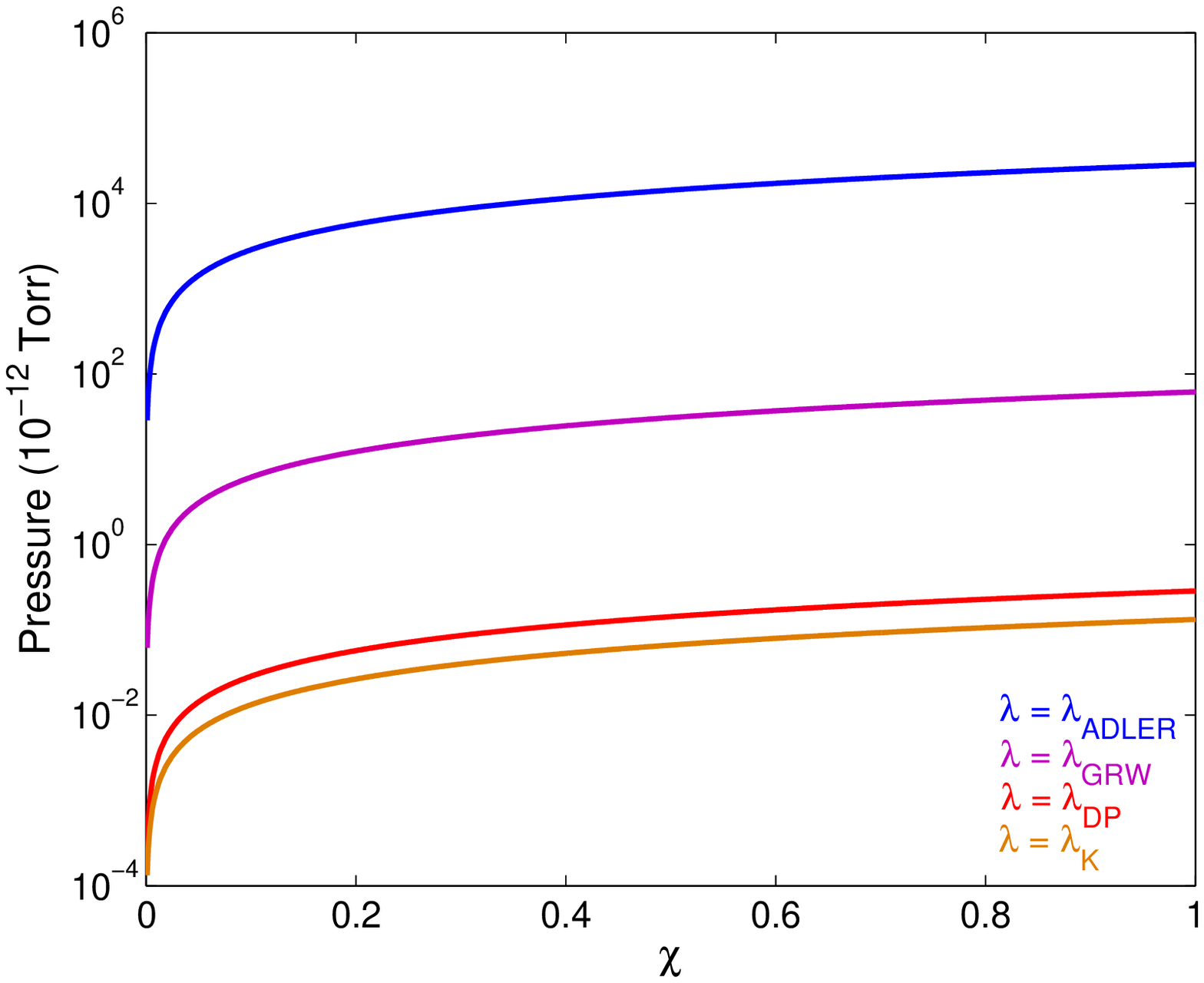}
\includegraphics[width=\columnwidth]{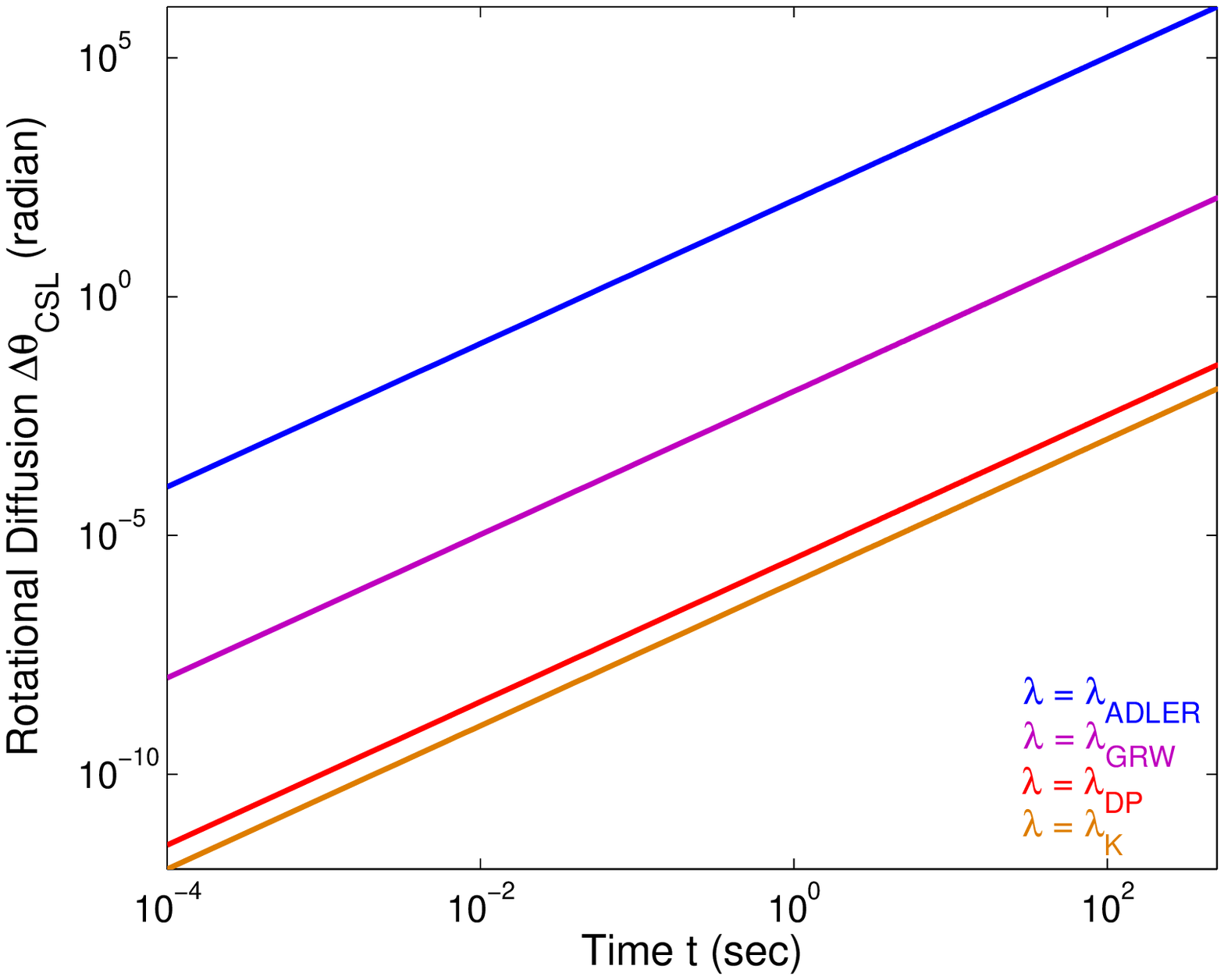}
\caption{\raggedright CSL diffusion of a Disc: Relative significance of thermal and CSL displacements constrains the  internal temperature $T_i$, plotted here  vs. the fraction $\epsilon = \Delta x_{RAD}/ \Delta x_{CSL}$ [top]; Relative significance of gas collisions and CSL displacement constrains required pressure  $P$ plotted here vs. the fraction $\chi=t_{CSL}/\tau_c$ [middle];  CSL rotational displacement $\Delta\theta_{CSL}$ vs. time for different models [bottom]. The different labelings on the rate constant $\lambda$ denote the different choices: ADLER, GRW,  and the two gravity models by Diosi-Penrose and Karolyhazy}

\end{figure}

\begin{table*}
\begin{tabular}{l l l l l l l}
\hline
$\Delta \theta_{CSL} (rad)$ \quad \quad & $t_{CSL}$ for $\lambda_{GRW}$ \quad \quad & $t_{CSL}$ for $\lambda_{ADLER}$ \quad \quad & $t_{Karolyhazy}$ \quad \quad \quad & $t_{DP}$ \quad \quad \quad \quad &$t_{QBD}$ \\ [1ex]
\hline 
$10^{-4}$ & $4.5\times 10^{-2}$ & $10^{-4}$ & 21 & 10 & 0.1 \\ 
$10^{-3}$ & 0.2 & $5\times 10^{-4}$ & 97 & 45 &  1 \\ 
$10^{-2}$ & 1 & $2\times 10^{-3}$ & 452 & 210 & 10\\ [1ex]
\hline 
\end{tabular}
\caption{Displacement time for Disc in $sec$, for different displacements and different models. The time for quantum Brownian motion $t_{QBD}$ exceeds $t_{CSL}$ and $t_{GRW}$ but is comparable or dominant over the displacement time in gravity models}
\end{table*}

Once again, we conclude that by keeping the CSL displacement low, at about a milli-radian, and demanding the CSL displacement time to be about a tenth of the normal Brownian displacement time, and assuming $\lambda$ to be $\lambda_{ADLER}$, the required pressure is about $10^{-10}$ Torr, which is achievable.

As for the gravity diffusion given by the K-model, by using $\lambda=\lambda_K$ and $T_e=1 K$, we obtain a very low internal temperature of about 1K and a very low pressure of about $10^{-4} pT$.

For the DP model we use Eqn $\eqref{disctemp}$ and Eqn $\eqref{discpressure}$  again with $T_e=1$ K and $\lambda_{DP}=10^{-23} sec^{-1}$. The estimates for internal temperature and pressure for the case of a disc using Di\'osi-Penrose are: $T_i \sim 1.18 K $, $ P \sim 1.33 \times 10^{-4}$ pT.

Figure 2 bottom panel shows the relative magnitudes of the CSL type random diffusion for different models. Clearly, gravity induced diffusion is considerably weaker than the CSL effect.

We observe from Table II that quantum Brownian displacements are comparable to or dominant over gravity diffusion, suggesting that even at such low pressures it may not be possible to unambiguously detect gravity induced random walk, unless we go to a higher mass and size, which lowers the required pressure even further.

\section{The case of a quantum oscillator}

We now address the question of how the CSL stochastic kicks could be looked for in the dynamics of an oscillator. According to Adler ~\cite{AdlerOsc}, the secular CSL induced increase in the energy of an oscillator is given as a function of time by
\begin{equation}
\delta E_{CSL} = \frac{\eta\hbar^{2}t}{2m}
\label{osctime}
\end{equation}
$\eta$ is the stochasticity parameter, which can be expressed as:
\begin{equation}
\eta_{GRW} = \frac{m}{m_0}\eta_0 = \eta_0 N
\end{equation}
[Equation. (2) of ~\cite{Bassi2:05}] where m$_0$ is the nucleon mass, $\eta_0 = \lambda_{GRW}/{{r_c}^{2}} = 10^{-6}$ cm$^{-2}$ sec$^{-1}$, and $N$ is the number of nucleons in the oscillator with localisation parameter $r_c$.
For the CSL model ~\cite{AdlerOsc},~\cite{RMP:2012}
\begin{subequations}
\begin{align}
\eta_{CSL} = \kappa N^{2/3} D^{4/3} \left(\pi {r_c}^{2}\right)^{-1/2}\\
\kappa = \frac{\lambda}{{(4\pi {r_c}^{2})}^{3/2}}
\end{align}
\end{subequations}
Here we assume the nucleon density, $D$ same as in ~\cite{AdlerOsc} i.e., $10^{24}$ cm$^{-3}$
Also, we see that the change in energy $\delta E$ is independent of the oscillator frequency. (A word about notation: we have switched from the symbol $\lambda$ to $\eta$, in order to be consistent with the notation used by Adler, so as to avoid confusion which could arise when the reader compares our equations with those in Adler's paper).

Now, from the partition function for a quantum harmonic oscillator, 
\begin{dmath}
Z = \exp\left[-(1/2)\beta \hbar \omega\right] \sum_{n = 0}^\infty (- n\beta\hbar\omega)\\
= \frac{\exp \left[-(1/2)\beta \hbar \omega\right]}{1 - \exp(-\beta \hbar \omega)}
\end{dmath}
where $\beta = 1/kT$, its mean energy is given by
\begin{equation}
\bar{E} = - \frac{\partial}{\partial\beta} ln Z \\
= \hbar\omega \left[\frac{1}{2} + \frac{1}{\exp(\beta\hbar\omega) - 1}\right]
\label{Emean}
\end{equation}

Considering as an example a time $t = 1$ sec, $\omega$ = 10 GHz and $\eta = \eta_{GRW}$, we obtain for a 10$^{12}$ nucleon system, from equation (\ref{osctime}), $\delta E_{CSL} \simeq 3.3 \times 10^{-37}$ ergs. For the same choice of parameters, the first term of Eqn. (\ref{Emean}) i.e. ${\hbar \omega}/{2}$ has a value of $\sim$ 10$^{-17}$ ergs, while the second (temperature-dependent) term is $\sim 10^{-50}$ ergs at temperature T = 1 mK. Hence, this term can be comfortably neglected for the purpose of comparison except at extremely high temperatures. 

[The situation for the oscillator should be contrasted with that for the rotating disc: in the latter case, the equivalent frequency for a quantum mechanical displacement (assuming no CSL effect) as given by 
$\omega = d\theta / dt$ is of the order of $10^{-3}$ Hz (see the discussion at the end of Sec. VI B in [CP]). In contrast, the frequency of 10 GHz for the quantum oscillator considered above is higher  than this by about fourteen orders of magnitude, which explains why the zero point energy of the quantum oscillator dominates over the CSL energy gain, unlike in the case of the disc.]

The above exercise shows that at temperatures close to zero, the internal energy dominates over the CSL gain, the former being about 18 orders higher in magnitude than the latter. Moreover, the internal energy of the oscillator increases as temperature goes higher, and thus it may be nearly impossible to detect the CSL energy gain of a quantum oscillator experimentally.

\begin{table}
\begin{tabular*}{\linewidth}{cccc}
\hline
Mass (amu) & \multicolumn{3}{c}{$\omega$ (Hz)}\\
    & $10^3$ & $10^6$ & $10^9$ \\
\hline               
$10^{6}$ & 1.6 $\times 10^{-9}$ & 6.5 $\times 10^{-7}$ & 4.1 $\times 10^{-4}$ \\ 
$10^{8}$ & 1.2 $\times 10^{-9}$ & 5.7 $\times 10^{-7}$ & 3.8 $\times 10^{-4}$ \\ 
$10^{10}$ & 9.6 $\times 10^{-10}$ & 5.1 $\times 10^{-7}$ & 3.5 $\times 10^{-4}$ \\ 
$10^{12}$ & 8 $\times 10^{-10}$ & 4.7 $\times 10^{-7}$ & 3.3 $\times 10^{-4}$ \\ [1ex]
\hline 
\end{tabular*}
\caption{Ambient Temperature (in K) for oscillator}
\end{table}

However, we also put forward the possibility that if an experimental setup could be devised such that the temperature-dependent term of the oscillator energy be solely measured, that is, the zero point energy background ${\hbar \omega}/{2}$ is subtracted, then one can detect the CSL induced energy gain of the oscillator at easily attainable temperatures.

In such a scenario, by requiring that the mean oscillator energy be a fraction $\epsilon$ of the CSL gain (\ref{osctime}), we get that
\begin{subequations}
\begin{align}
\label{fracEq}
\hbar\omega \left[\frac{1}{\exp(\beta\hbar\omega) - 1}\right] = \epsilon \frac{\eta\hbar^{2}t}{2m} \\
T = \frac{\hbar \omega/k}{\ln\left(\frac{2m\omega}{\epsilon\eta\hbar t} + 1\right)}
\end{align}
\end{subequations}

For $t = 1$ sec, $\omega$ = 10 GHz as before, $\epsilon$ = 0.1 and $\eta = \eta_{GRW}$, we obtain the desired temperature to be $\sim$ 2 mK. With $\eta_{CSL}$, the value turns out to be 3 mK. However here, due to measurement uncertainties, we must take care that the CSL energy observations should be done for time intervals that are greater than the experimental resolution, that is 
\begin{equation*}
t_{CSL} \geq \frac{Q}{\omega}
\end{equation*}
where $Q$ is the quality factor of the oscillator. With the previous value of $\omega$ and Q = 10$^{5}$, $t_{CSL}$ should be $\geq 10^{-5}$ sec.

Next, in order to estimate the pressure $P$ of the ambient medium, we require that the time  $t_{CSL}$
over which we observe the secular heating, should be a fraction $\chi$ of the time between two collisions of an ambient molecule with the oscillator (assumed to be a plate with area $A$).

\begin{table}
\begin{tabular*}{\linewidth}{cccc}
\hline
Mass (amu) & \multicolumn{3}{c}{$\omega$ (Hz)}\\
    & $10^3$ & $10^6$ & $10^9$ \\
    \hline
$10^{6}$ & 3 $\times 10^{-4}$ & 0.0063 & 0.1576 \\ 
$10^{8}$ & 2.69 $\times 10^{-4}$ & 0.0059 & 0.1515 \\ 
$10^{10}$ & 2.42 $\times 10^{-4}$ & 0.0056 & 0.1461 \\ 
$10^{12}$ & 2.21 $\times 10^{-4}$ & 0.0053 & 0.1412 \\ [1ex]
\hline 
\end{tabular*}
\caption{Ambient Pressure (in pico Torr) for oscillator}
Values used: $\epsilon$ = 0.1, $\chi$ = 0.1, time of evolution $t$ = 1 sec, density $D$ = 10$^{24}$ g/cc, $\lambda = \lambda_{GRW} = 10^{-16}$ sec$^{-1}$
\end{table}

Using Eqns. (\ref{osctime}) and (\ref{fracEq}) to compare with the time $\tau_c$ between collisions
\begin{equation}
\tau_c = \frac{1.3 \times 10^{-9}}{A} \left(T/T_0\right)^{1/2} (P pT)^{-1}
\end{equation}
we obtain the relation
\begin{dmath}
P  \; pT = \chi \epsilon\frac{1.3 \times 10^{-9}\eta \hbar }{2m\omega A} \left(T/T_0\right)^{1/2}\times
\left[\exp\left(\beta\hbar\omega\right) - 1\right]
\end{dmath}
So, for Adler's representative choice of $10^{12}$ nucleons, $\eta_{GRW}$ = $10^{6}$ cm$^{-2}$ 
sec$^{-1}$, plate area $10^{-12}$ cm$^{2}$, $\omega = 10$ GHz, $\epsilon$ = 0.1, $\chi = 0.1$ and $T = 1.6 \times 10^{-3}$ K, the pressure comes out to be $\sim$ 0.3 pico Torr. Because the pressure now depends linearly on $\eta$, we see the dramatic result that if the value of the stochastic parameter $\lambda$ were to be raised to the value $\lambda_{ADLER}$ (higher by some eight orders of magnitude) the required pressure would only be about $10^{-5}$ Torr. The CSL energy gain would be about $10^{-29}$ ergs in 1 sec. In classical terms this corresponds to a displacement of about $10^{-18}$ cm.

Tables III and IV show the required temperature and pressure as a function of oscillator mass and frequency. It is clear that higher mass and higher frequency are ideal from the point of view of achievable pressures and temperatures. On the other hand, the CSL energy gain and the positional displacement will be higher for lower masses. The CSL energy gain rises inversely as mass, and the positional displacement also rises inversely as the square root of mass. Thus for a 10 GHz oscillator of a million amu, the positional displacement would be about $10^{-15}$ cm, and the required pressure and temperature are roughly in the 
pico-Torr and micro-Kelvin range respectively.

In the next section we now discuss how the required internal and external temperatures, and the low pressures, can be attained in laboratory experiments, using currently available technology. In so doing, we conclude that detection of random diffusion constitutes an achievable test for confirmation / refutation of the GRW and CSL models.

\section{Consideration of experimental realisations}
Here we discuss the possibility to perform anomalous Brownian motion tests with existing technology and find quite plausible solutions. From the theoretical estimates performed above, it is clear that we need relatively large particles (of various shapes). Interestingly, experiments favoured in recent years to test collapse models require the generation of a spatial superposition state of a massive object~\cite{hornberger2012colloquium, eibenberger2013matter}, realising a quantum state of clear Macroscpicity~\cite{nimmrichter2013macroscopicity}. CSL models then predict that this quantum superposition will be destroyed by a yet unknown mechanism of collapse. Here we have investigated a different effect, which is a heating effect and its detection requires us to avoid other heating effects from  dominating. The latter effect (i.e. CSL heating) appears to be easier to be realised in experiments with existing technology, as we shall discuss in some detail now. A similar and related experiment also targeting the CSL heating effect has been proposed recently by two independent studies~\cite{bahrami2014testing, Bahrami2014, nimmrichter2014optomechanical}, as well as by  heating of an atomic Bose Einstein Condensate~\cite{Pearle2014}. So we expect more experimental possibilities to open up, if more and more experimentalists get interested in the test of CSL models and similar effects.

\subsection{General conditions:} The requirements on {\it pressure} are achievable in ultra-high vacuum experiments such as usual in surface science or cold atom experiments where pressures of 10$^{-12}$ Torr are routinely achieved. It requires a procedure called the `bake out' to achieve a vacuum below 10$^{-9}$ torr. The whole vacuum chamber with all internal parts has to be heated to temperatures of around 150 degree Celsius for one to two weeks. While this requires a careful selection of materials and vacuum components, it is a standard procedure and a lot of relevant knowledge exists. Record low pressures of 5$\cdot$10$^{-17}$ Torr have been reported~\cite{gabrielse1990thousandfold}, but are not achievable in common experimental configurations.

The {\it temperature} requirements represent a more severe constraint to the experiment. Temperatures in the range below room temperatures for both internal and external degrees of freedom are needed, depending on the specific model under test. In general, cryogenic technology is commercially available down to some 1 mK in dilution cryostats. However such cooling techniques, while affecting all degrees of freedom of a particle/structure, require direct thermal contact of the body to cool with the cold finger of the cryostat or to rely on very slow thermal radiation exchange between the cold cryostat and body to cool, if levitated. The latter setting results in an extremely small cooling rate, which is not practical.  

It is clear that test of the CSL heating effect, at least for the GRW parameters, requires cooling of the levitated object. On the first hand, predominantly the {\it external}, centre-of-mass motion has to be cooled to make the CSL Brownian motion a visible effect. Luckily, recent progress in experiments has shown optical cooling to about 10 mK of the centre of mass motion for 100 nm sized particles. Lowest temperatures so far have been achieved by parametric feedback cooling, where the position of the particle is optically tracked and the intensity of the trapping laser is modified accordingly~\cite{li2011millikelvin, gieseler2012subkelvin}. Feedback cooling can be done to affect all three degrees of centre-of-mass motion. A somewhat different technique is optical cavity cooling, where the centre of mass motion is coupled to the light mode in the cavity and the natural cavity decay therefore also cools the particle's motion. Three different experimental configurations have shown the proof of principle of cavity cooling in one dimension of the motion~\cite{kiesel2013cavity, asenbaum2013cavity, millen2014optomechanical}. Now the cooling rate has to be improved and the other two degrees have to be affected as well. For the test of CSL heating, the achieved 10 mK are already sufficient { (see Figs. 1 and 2 above)}.

The bigger problem is the cooling of {\it internal} degrees of freedom of levitated particles; by internal degrees of freedom we mean vibrations, rotations and electronic excitations.There has been no demonstration of any technique so far. Under vacuum, internal and external degrees of freedom are not coupled. So cooling the external degrees of freedom has no significant effect on the internal temperature. However, there are some promising first ideas, which have been proposed, such as cavity cooling which links to internal degrees of freedom~\cite{morigi2007cavity}  or the so-called Raman cooling of solids with a specific internal structure, so that the optical field can directly extract phonons from the internal thermal distribution.This technique might be applicable to nanoparticles. Experiments with particles on substrates have been performed already~\cite{seletskiy2010laser}. Cooling the internal degrees of freedom is the biggest experimental challenge remaining. However from our estimates  with  equations ({\ref{Tintsph}) and (\ref{Tintdisc}) we seem to be able to perform the test of CSL heating with Adler parameters without internal state cooling. 

\subsection{The choice of the system:} 
In technical terms, we need a mechanical harmonic oscillator with a high $Q$ factor, which means that a once excited oscillation goes on for a long time without damping, in other words without  heating from an external heat bath. For such an undamped oscillator we expect a narrow spectral line associated with the mechanical oscillation: $Q =\omega / \delta\omega$, with $\delta\omega$ being the width of the spectral peak. We want to avoid any external heating, which is larger than the incredibly small CSL heating effect. That makes the experiment challenging, but it seems feasible with existing technology. 

This means that all methods with suspended objects, even if the suspension is as small as a nanowire, do not seem too promising. The Q factor for the mechanical oscillation for suspended devices can hardly be better than 10$^6$  which seems to rule out mechanically clamped systems, like micro- or nano fabricated cantilever structures. However we do not want to exclude this possibility in general as success maybe possible with very sophisticated structures such as for instance phononic crystal structures~\cite{chan2011laser}, which reduce thermal dissipation to a minimum. In general, quantum optomechanical systems show very promising features for tests of non-classical behaviour at mesoscopic scales, but mostly are realised in clamped geometries~\cite{aspelmeyer2013cavity}. A more natural choice seems to be levitated objects, which do not have any clamping losses nor any dissipation through mechanical links.

There are plenty of {\it noise effects} which could easily heat the system under consideration more substantially than the desired CSL heating effect. Such noise effects include vibrations generated by the experimental environment by vacuum pumps and compressors, electrical read out noise for detectors such as photo-diodes or Squid position sensors, intensity and frequency noise of the trapping laser (in case optical levitation is needed), heating by absorption of trapped laser photons by the particle, which will heat up the internal temperature etc. For each experimental configuration such systematic effects have to be carefully checked. This again is an argument for experiments with levitated particles, but also here a vibration of the trap itself (which will be realised by lenses, mirrors or magnetic coils) will need critical assessment and carefully engineered solutions. All systematic noise effects can be quantified by their effect on the $Q$-factor or on external and internal temperature. 

From our analysis above we see that the particular {\it shape of the particle}  is important for the observation of the CSL heating/diffusion effect. The disc geometry gives more promising results. A technical difficulty, which we would like to mention here, is that trapping and levitating a shape different from the sphere has to be thought in careful details. For instance the trapping of a disc, which is free to rotate around the axis crossing the flat diameter of the disc, is quite tricky. In general, Earnshaw's theorem has to be fulfilled for stable trapping and to build a trap for a sphere is straightforward, while the trapping of asymmetric shapes such as a disc is more advanced. Somehow, the symmetry of the trapping field has to be designed for the symmetry of the particle to be trapped. Also for each shape of the particle chosen the CSL heating effect has to be recalculated, which can be difficult in detail. We mention this, to emphasize that such experimental details can be non-trivial.

{\bf Optical levitation} or tweezing has a long and successful history with plentiful applications in physics, chemistry and life sciences~\cite{ashkin2000history}. The majority of such experiments are performed with particles in solution. Recently and as already mentioned above different cooling/stabilisation techniques have been demonstrated in vacuum as small as $10^{-6}$ mbar. The Q factor predicted is as high as $10^{12}$ at ultra-high vacuum, but needs to be shown by experiment~\cite{gieseler2013thermal}.

One limiting effect, which is expected to reduce $Q$ is the absorption of photons from the trapping laser. This effect can be reduced with the right choice of material of the particle. At the moment the lowest absorption cross section is predicted for a silicon nano particle in a laser dipole trap at 1550 nm~\cite{asenbaum2013cavity, bateman2013near}. While this heating process can be minimised, there will always be heating of the internal temperature, which would limit ability to observe the CSL heating in the experiment.  

This means the trapping and cooling of the external centre of mass motion are possible, but the effects, intrinsic to the optical levitation, of heating by absorption and the cooling of internal states is still problematic. 

{\bf Magnetic levitation:} 
To overcome this absorption heating problem magnetic trapping at very low temperatures has been proposed \cite{Romero2012magneto}. Magnetic levitation has a long history as well for ultra-precise sensing and metrology, including gravitational effects \cite{Coccia:1998}. Such magnetic levitation set-ups seem to have many favourable properties for the test of small effects such as the CSL heating effect. The low temperatures will also help to reach very low pressures, as the cold surfaces of the solid parts of the experiment will adsorb atoms and molecules, which is known as cryogenic pump.  {\bf Ion trap experiments} are also  promising for CSL tests, however it will be challenging to build a stable Paul or Penning ion trap, where the noise in the trapping field is smaller than the CSL heating effect. That seems challenging. For instance for a Paul trap, micro motion effects will have to be balanced almost perfectly or separated for the CSL effect in the frequency domain.  In general, magnetic levitation seems to be seen as the method with the lowest systematic noise generated by the trap 
\cite{Brandt:1989}, which would be the preferred way to perform CSL tests.    

While the test of gravity induced collapse models is clearly out of reach for experiments with levitated particles, the test of CSL effect with Adler as well as with GRW parameters seems feasible. 

\section{Conclusions}
 
It is important to emphasise that while quantum theory is extremely successful and not contradicted by any experiment, there is a vast range in the parameter space (objects of masses ranging from $10^{5}$ amu to $10^{18}$ amu), over which the theory has not been tested. Precision tests of quantum theory in this range are of great importance, from the point of view of confirming the theory, or finding new effects. The theory of CSL provides a significant benchmark against which quantum theory can be tested, because it makes experimental predictions which are quantitatively different from that of quantum theory. This in itself, apart from testing CSL  as a possible explanation for collapse of the wave function, is an important motivation for carrying out experiments of the kind suggested in this paper.

During the last fifteen years or so, experiments to test quantum theory in the mesoscopic regime have been picking up momentum, especially as new ideas have been put forth with regard to quantum systems and techniques to be used. The lead provided by breakthrough experiments  and subsequent advances in molecular interferometry has been supplemented by significant advances in optomechanics and cooling of optomechanical devices. Progress has been made towards generating Schr\"odinger cat states for ever larger systems, with increasing position separation amongst the superposed states. In addition, new ideas for testing CSL have been proposed, such as spectral line broadening, and constraints coming from heating of atomic BECs.

Our proposal here, based on a re-analysis of the earlier work of Collett and Pearle, suggests to look for the effect of CSL heating in the random diffusion of the affected mesoscopic object, here assumed to have a size of 100 nm = 10$^{-5}$ cm, and hence a mass of about $10^{-15}$ gms. This set of parameter values appears to provide temperature and pressure requirements which are achievable with current technology. We thus hope that our work will encourage experimentalists to set up experiments to look for CSL diffusion. In combination with experiments in matter wave interferometry, optomechanics, and frequency domain tests of spectral line broadening, tests of random diffusion could serve to put stronger bounds on departure from quantum theory, in the near future.

\section*{}{\bf Acknowledgements} 
We are grateful to Philip Pearle for helpful comments on a previous version of this manuscript.
We acknowledge help from Marco Moreno during early stages of this work. We wish to thank the UK funding agency EP- SRC for support under grant (EP/J014664/1), the Foundational Questions Institute (FQXi), and the John F Templeton foundation under grant (39530).


\bibliography{biblioqmts3}

\def\polhk#1{\setbox0=\hbox{#1}{\ooalign{\hidewidth
  \lower1.5ex\hbox{`}\hidewidth\crcr\unhbox0}}} \def\cprime{$'$}
  \def\cprime{$'$}
\begin{thebibliography}{40}%
\makeatletter
\providecommand \@ifxundefined [1]{%
 \@ifx{#1\undefined}
}%
\providecommand \@ifnum [1]{%
 \ifnum #1\expandafter \@firstoftwo
 \else \expandafter \@secondoftwo
 \fi
}%
\providecommand \@ifx [1]{%
 \ifx #1\expandafter \@firstoftwo
 \else \expandafter \@secondoftwo
 \fi
}%
\providecommand \natexlab [1]{#1}%
\providecommand \enquote  [1]{``#1''}%
\providecommand \bibnamefont  [1]{#1}%
\providecommand \bibfnamefont [1]{#1}%
\providecommand \citenamefont [1]{#1}%
\providecommand \href@noop [0]{\@secondoftwo}%
\providecommand \href [0]{\begingroup \@sanitize@url \@href}%
\providecommand \@href[1]{\@@startlink{#1}\@@href}%
\providecommand \@@href[1]{\endgroup#1\@@endlink}%
\providecommand \@sanitize@url [0]{\catcode `\\12\catcode `\$12\catcode
  `\&12\catcode `\#12\catcode `\^12\catcode `\_12\catcode `\%12\relax}%
\providecommand \@@startlink[1]{}%
\providecommand \@@endlink[0]{}%
\providecommand \url  [0]{\begingroup\@sanitize@url \@url }%
\providecommand \@url [1]{\endgroup\@href {#1}{\urlprefix }}%
\providecommand \urlprefix  [0]{URL }%
\providecommand \Eprint [0]{\href }%
\providecommand \doibase [0]{http://dx.doi.org/}%
\providecommand \selectlanguage [0]{\@gobble}%
\providecommand \bibinfo  [0]{\@secondoftwo}%
\providecommand \bibfield  [0]{\@secondoftwo}%
\providecommand \translation [1]{[#1]}%
\providecommand \BibitemOpen [0]{}%
\providecommand \bibitemStop [0]{}%
\providecommand \bibitemNoStop [0]{.\EOS\space}%
\providecommand \EOS [0]{\spacefactor3000\relax}%
\providecommand \BibitemShut  [1]{\csname bibitem#1\endcsname}%
\let\auto@bib@innerbib\@empty
\bibitem [{\citenamefont {Pearle}(1989)}]{PEARLE2}%
  \BibitemOpen
  \bibfield  {author} {\bibinfo {author} {\bibfnamefont {P.}~\bibnamefont
  {Pearle}},\ }\href@noop {} {\bibfield  {journal} {\bibinfo  {journal} {Phys.
  Rev. A}\ }\textbf {\bibinfo {volume} {39}},\ \bibinfo {pages} {2277}
  (\bibinfo {year} {1989})}\BibitemShut {NoStop}%
\bibitem [{\citenamefont {Ghirardi}\ \emph {et~al.}(1990)\citenamefont
  {Ghirardi}, \citenamefont {Pearle},\ and\ \citenamefont
  {Rimini}}]{Ghirardi2:90}%
  \BibitemOpen
  \bibfield  {author} {\bibinfo {author} {\bibfnamefont {G.~C.}\ \bibnamefont
  {Ghirardi}}, \bibinfo {author} {\bibfnamefont {P.}~\bibnamefont {Pearle}}, \
  and\ \bibinfo {author} {\bibfnamefont {A.}~\bibnamefont {Rimini}},\
  }\href@noop {} {\bibfield  {journal} {\bibinfo  {journal} {Phys. Rev. A}\
  }\textbf {\bibinfo {volume} {42}},\ \bibinfo {pages} {78} (\bibinfo {year}
  {1990})}\BibitemShut {NoStop}%
\bibitem [{\citenamefont {Bassi}\ and\ \citenamefont
  {Ghirardi}(2003)}]{Bassi:03}%
  \BibitemOpen
  \bibfield  {author} {\bibinfo {author} {\bibfnamefont {A.}~\bibnamefont
  {Bassi}}\ and\ \bibinfo {author} {\bibfnamefont {G.~C.}\ \bibnamefont
  {Ghirardi}},\ }\href@noop {} {\bibfield  {journal} {\bibinfo  {journal}
  {Phys. Rep.}\ }\textbf {\bibinfo {volume} {379}},\ \bibinfo {pages} {257}
  (\bibinfo {year} {2003})}\BibitemShut {NoStop}%
\bibitem [{\citenamefont {Bassi}\ \emph {et~al.}(2013)\citenamefont {Bassi},
  \citenamefont {Lochan}, \citenamefont {Satin}, \citenamefont {Singh},\ and\
  \citenamefont {Ulbricht}}]{RMP:2012}%
  \BibitemOpen
  \bibfield  {author} {\bibinfo {author} {\bibfnamefont {A.}~\bibnamefont
  {Bassi}}, \bibinfo {author} {\bibfnamefont {K.}~\bibnamefont {Lochan}},
  \bibinfo {author} {\bibfnamefont {S.}~\bibnamefont {Satin}}, \bibinfo
  {author} {\bibfnamefont {T.~P.}\ \bibnamefont {Singh}}, \ and\ \bibinfo
  {author} {\bibfnamefont {H.}~\bibnamefont {Ulbricht}},\ }\href@noop {}
  {\bibfield  {journal} {\bibinfo  {journal} {Rev. Mod. Phys.}\ }\textbf
  {\bibinfo {volume} {85}},\ \bibinfo {pages} {471} (\bibinfo {year}
  {2013})}\BibitemShut {NoStop}%
\bibitem [{\citenamefont {Bahrami}\ \emph
  {et~al.}(2014{\natexlab{a}})\citenamefont {Bahrami}, \citenamefont
  {Paternostro}, \citenamefont {Bassi},\ and\ \citenamefont
  {Ulbricht}}]{Bahrami2014}%
  \BibitemOpen
  \bibfield  {author} {\bibinfo {author} {\bibfnamefont {M.}~\bibnamefont
  {Bahrami}}, \bibinfo {author} {\bibfnamefont {M.}~\bibnamefont
  {Paternostro}}, \bibinfo {author} {\bibfnamefont {A.}~\bibnamefont {Bassi}},
  \ and\ \bibinfo {author} {\bibfnamefont {H.}~\bibnamefont {Ulbricht}},\
  }\href@noop {} {\bibfield  {journal} {\bibinfo  {journal} {Phys. Rev. Lett.}\
  }\textbf {\bibinfo {volume} {112}},\ \bibinfo {pages} {210404} (\bibinfo
  {year} {2014}{\natexlab{a}})}\BibitemShut {NoStop}%
\bibitem [{\citenamefont {Lal\"oe}\ \emph {et~al.}(2014)\citenamefont
  {Lal\"oe}, \citenamefont {Mullin},\ and\ \citenamefont
  {Pearle}}]{Pearle2014}%
  \BibitemOpen
  \bibfield  {author} {\bibinfo {author} {\bibfnamefont {F.}~\bibnamefont
  {Lal\"oe}}, \bibinfo {author} {\bibfnamefont {W.~J.}\ \bibnamefont {Mullin}},
  \ and\ \bibinfo {author} {\bibfnamefont {P.}~\bibnamefont {Pearle}},\
  }\href@noop {} {\ \textbf {\bibinfo {volume} {arXiv:1409.5388}} (\bibinfo
  {year} {2014})}\BibitemShut {NoStop}%
\bibitem [{\citenamefont {Ghirardi}\ \emph {et~al.}(1986)\citenamefont
  {Ghirardi}, \citenamefont {Rimini},\ and\ \citenamefont
  {Weber}}]{Ghirardi:86}%
  \BibitemOpen
  \bibfield  {author} {\bibinfo {author} {\bibfnamefont {G.~C.}\ \bibnamefont
  {Ghirardi}}, \bibinfo {author} {\bibfnamefont {A.}~\bibnamefont {Rimini}}, \
  and\ \bibinfo {author} {\bibfnamefont {T.}~\bibnamefont {Weber}},\
  }\href@noop {} {\bibfield  {journal} {\bibinfo  {journal} {Phys. Rev. D}\
  }\textbf {\bibinfo {volume} {34}},\ \bibinfo {pages} {470} (\bibinfo {year}
  {1986})}\BibitemShut {NoStop}%
\bibitem [{\citenamefont {Pearle}(2014)}]{Pearlepvt}%
  \BibitemOpen
  \bibfield  {author} {\bibinfo {author} {\bibfnamefont {P.}~\bibnamefont
  {Pearle}},\ }\href@noop {} {\ \textbf {\bibinfo {volume} {Private
  Communication}} (\bibinfo {year} {2014})}\BibitemShut {NoStop}%
\bibitem [{\citenamefont {Collett}\ and\ \citenamefont
  {Pearle}(2003)}]{PEARLE5}%
  \BibitemOpen
  \bibfield  {author} {\bibinfo {author} {\bibfnamefont {B.}~\bibnamefont
  {Collett}}\ and\ \bibinfo {author} {\bibfnamefont {P.}~\bibnamefont
  {Pearle}},\ }\href@noop {} {\bibfield  {journal} {\bibinfo  {journal} {Found.
  Phys.}\ }\textbf {\bibinfo {volume} {33}},\ \bibinfo {pages} {1495} (\bibinfo
  {year} {2003})}\BibitemShut {NoStop}%
\bibitem [{\citenamefont {Adler}(2007)}]{Adler3:07}%
  \BibitemOpen
  \bibfield  {author} {\bibinfo {author} {\bibfnamefont {S.~L.}\ \bibnamefont
  {Adler}},\ }\href@noop {} {\bibfield  {journal} {\bibinfo  {journal} {J.
  Phys. A}\ }\textbf {\bibinfo {volume} {40}},\ \bibinfo {pages} {2935}
  (\bibinfo {year} {2007})}\BibitemShut {NoStop}%
\bibitem [{\citenamefont {Karolyhazy}(1966)}]{Karolyhazi:66}%
  \BibitemOpen
  \bibfield  {author} {\bibinfo {author} {\bibfnamefont {F.}~\bibnamefont
  {Karolyhazy}},\ }\href@noop {} {\bibfield  {journal} {\bibinfo  {journal}
  {Nuovo Cimento}\ }\textbf {\bibinfo {volume} {42A}},\ \bibinfo {pages} {390}
  (\bibinfo {year} {1966})}\BibitemShut {NoStop}%
\bibitem [{\citenamefont {Karolyhazy}\ \emph {et~al.}(1986)\citenamefont
  {Karolyhazy}, \citenamefont {Frenkel},\ and\ \citenamefont
  {Luk\'acs}}]{Karolyhazi:86}%
  \BibitemOpen
  \bibfield  {author} {\bibinfo {author} {\bibfnamefont {F.}~\bibnamefont
  {Karolyhazy}}, \bibinfo {author} {\bibfnamefont {A.}~\bibnamefont {Frenkel}},
  \ and\ \bibinfo {author} {\bibfnamefont {B.}~\bibnamefont {Luk\'acs}},\ }in\
  \href@noop {} {\emph {\bibinfo {booktitle} {Quantum concepts in space and
  time}}},\ \bibinfo {editor} {edited by\ \bibinfo {editor} {\bibfnamefont
  {R.}~\bibnamefont {Penrose}}\ and\ \bibinfo {editor} {\bibfnamefont {C.~J.}\
  \bibnamefont {Isham}}}\ (\bibinfo  {publisher} {Clarendon},\ \bibinfo
  {address} {Oxford},\ \bibinfo {year} {1986})\BibitemShut {NoStop}%
\bibitem [{\citenamefont {Frenkel}(1990)}]{Frenkel:90}%
  \BibitemOpen
  \bibfield  {author} {\bibinfo {author} {\bibfnamefont {A.}~\bibnamefont
  {Frenkel}},\ }\href@noop {} {\bibfield  {journal} {\bibinfo  {journal}
  {Found. Phys.}\ }\textbf {\bibinfo {volume} {20}},\ \bibinfo {pages} {159}
  (\bibinfo {year} {1990})}\BibitemShut {NoStop}%
\bibitem [{\citenamefont {Di{\'o}si}(1987)}]{Diosi:87}%
  \BibitemOpen
  \bibfield  {author} {\bibinfo {author} {\bibfnamefont {L.}~\bibnamefont
  {Di{\'o}si}},\ }\href {\doibase DOI: 10.1016/0375-9601(87)90681-5} {\bibfield
   {journal} {\bibinfo  {journal} {Physics Letters A}\ }\textbf {\bibinfo
  {volume} {120}},\ \bibinfo {pages} {377 } (\bibinfo {year}
  {1987})}\BibitemShut {NoStop}%
\bibitem [{\citenamefont {Di\'osi}(1989)}]{Diosi:89}%
  \BibitemOpen
  \bibfield  {author} {\bibinfo {author} {\bibfnamefont {L.}~\bibnamefont
  {Di\'osi}},\ }\href {\doibase 10.1103/PhysRevA.40.1165} {\bibfield  {journal}
  {\bibinfo  {journal} {Phys. Rev. A}\ }\textbf {\bibinfo {volume} {40}},\
  \bibinfo {pages} {1165} (\bibinfo {year} {1989})}\BibitemShut {NoStop}%
\bibitem [{\citenamefont {Bahrami}\ \emph
  {et~al.}(2014{\natexlab{b}})\citenamefont {Bahrami}, \citenamefont {Smirne},\
  and\ \citenamefont {Bassi}}]{Bassi2014}%
  \BibitemOpen
  \bibfield  {author} {\bibinfo {author} {\bibfnamefont {M.}~\bibnamefont
  {Bahrami}}, \bibinfo {author} {\bibfnamefont {A.}~\bibnamefont {Smirne}}, \
  and\ \bibinfo {author} {\bibfnamefont {A.}~\bibnamefont {Bassi}},\
  }\href@noop {} {\ \textbf {\bibinfo {volume} {arXiv:1408.6460}} (\bibinfo
  {year} {2014}{\natexlab{b}})}\BibitemShut {NoStop}%
\bibitem [{\citenamefont {Adler}(2005)}]{AdlerOsc}%
  \BibitemOpen
  \bibfield  {author} {\bibinfo {author} {\bibfnamefont {S.~L.}\ \bibnamefont
  {Adler}},\ }\href@noop {} {\bibfield  {journal} {\bibinfo  {journal} {J.
  Phys. A}\ }\textbf {\bibinfo {volume} {38}},\ \bibinfo {pages} {2729}
  (\bibinfo {year} {2005})}\BibitemShut {NoStop}%
\bibitem [{\citenamefont {Romero-Isart}(2011)}]{Romero2012decoherence}%
  \BibitemOpen
  \bibfield  {author} {\bibinfo {author} {\bibfnamefont {O.}~\bibnamefont
  {Romero-Isart}},\ }\href@noop {} {\bibfield  {journal} {\bibinfo  {journal}
  {Phys. Rev. A}\ }\textbf {\bibinfo {volume} {84}},\ \bibinfo {pages} {052121}
  (\bibinfo {year} {2011})}\BibitemShut {NoStop}%
\bibitem [{\citenamefont {Bassi}(2005)}]{Bassi2:05}%
  \BibitemOpen
  \bibfield  {author} {\bibinfo {author} {\bibfnamefont {A.}~\bibnamefont
  {Bassi}},\ }\href@noop {} {\bibfield  {journal} {\bibinfo  {journal} {J.
  Phys. A}\ }\textbf {\bibinfo {volume} {38}},\ \bibinfo {pages} {3173}
  (\bibinfo {year} {2005})}\BibitemShut {NoStop}%
\bibitem [{\citenamefont {Hornberger}\ \emph {et~al.}(2012)\citenamefont
  {Hornberger}, \citenamefont {Gerlich}, \citenamefont {Haslinger},
  \citenamefont {Nimmrichter},\ and\ \citenamefont
  {Arndt}}]{hornberger2012colloquium}%
  \BibitemOpen
  \bibfield  {author} {\bibinfo {author} {\bibfnamefont {K.}~\bibnamefont
  {Hornberger}}, \bibinfo {author} {\bibfnamefont {S.}~\bibnamefont {Gerlich}},
  \bibinfo {author} {\bibfnamefont {P.}~\bibnamefont {Haslinger}}, \bibinfo
  {author} {\bibfnamefont {S.}~\bibnamefont {Nimmrichter}}, \ and\ \bibinfo
  {author} {\bibfnamefont {M.}~\bibnamefont {Arndt}},\ }\href@noop {}
  {\bibfield  {journal} {\bibinfo  {journal} {Reviews of Modern Physics}\
  }\textbf {\bibinfo {volume} {84}},\ \bibinfo {pages} {157} (\bibinfo {year}
  {2012})}\BibitemShut {NoStop}%
\bibitem [{\citenamefont {Eibenberger}\ \emph {et~al.}(2013)\citenamefont
  {Eibenberger}, \citenamefont {Gerlich}, \citenamefont {Arndt}, \citenamefont
  {Mayor},\ and\ \citenamefont {T{\"u}xen}}]{eibenberger2013matter}%
  \BibitemOpen
  \bibfield  {author} {\bibinfo {author} {\bibfnamefont {S.}~\bibnamefont
  {Eibenberger}}, \bibinfo {author} {\bibfnamefont {S.}~\bibnamefont
  {Gerlich}}, \bibinfo {author} {\bibfnamefont {M.}~\bibnamefont {Arndt}},
  \bibinfo {author} {\bibfnamefont {M.}~\bibnamefont {Mayor}}, \ and\ \bibinfo
  {author} {\bibfnamefont {J.}~\bibnamefont {T{\"u}xen}},\ }\href@noop {}
  {\bibfield  {journal} {\bibinfo  {journal} {Physical Chemistry Chemical
  Physics}\ }\textbf {\bibinfo {volume} {15}},\ \bibinfo {pages} {14696}
  (\bibinfo {year} {2013})}\BibitemShut {NoStop}%
\bibitem [{\citenamefont {Nimmrichter}\ and\ \citenamefont
  {Hornberger}(2013)}]{nimmrichter2013macroscopicity}%
  \BibitemOpen
  \bibfield  {author} {\bibinfo {author} {\bibfnamefont {S.}~\bibnamefont
  {Nimmrichter}}\ and\ \bibinfo {author} {\bibfnamefont {K.}~\bibnamefont
  {Hornberger}},\ }\href@noop {} {\bibfield  {journal} {\bibinfo  {journal}
  {Physical review letters}\ }\textbf {\bibinfo {volume} {110}},\ \bibinfo
  {pages} {160403} (\bibinfo {year} {2013})}\BibitemShut {NoStop}%
\bibitem [{\citenamefont {Bahrami}\ \emph
  {et~al.}(2014{\natexlab{c}})\citenamefont {Bahrami}, \citenamefont {Bassi},\
  and\ \citenamefont {Ulbricht}}]{bahrami2014testing}%
  \BibitemOpen
  \bibfield  {author} {\bibinfo {author} {\bibfnamefont {M.}~\bibnamefont
  {Bahrami}}, \bibinfo {author} {\bibfnamefont {A.}~\bibnamefont {Bassi}}, \
  and\ \bibinfo {author} {\bibfnamefont {H.}~\bibnamefont {Ulbricht}},\
  }\href@noop {} {\bibfield  {journal} {\bibinfo  {journal} {Physical Review
  A}\ }\textbf {\bibinfo {volume} {89}},\ \bibinfo {pages} {032127} (\bibinfo
  {year} {2014}{\natexlab{c}})}\BibitemShut {NoStop}%
\bibitem [{\citenamefont {Nimmrichter}\ \emph {et~al.}(2014)\citenamefont
  {Nimmrichter}, \citenamefont {Hornberger},\ and\ \citenamefont
  {Hammerer}}]{nimmrichter2014optomechanical}%
  \BibitemOpen
  \bibfield  {author} {\bibinfo {author} {\bibfnamefont {S.}~\bibnamefont
  {Nimmrichter}}, \bibinfo {author} {\bibfnamefont {K.}~\bibnamefont
  {Hornberger}}, \ and\ \bibinfo {author} {\bibfnamefont {K.}~\bibnamefont
  {Hammerer}},\ }\href@noop {} {\bibfield  {journal} {\bibinfo  {journal}
  {arXiv preprint arXiv:1405.2868}\ } (\bibinfo {year} {2014})}\BibitemShut
  {NoStop}%
\bibitem [{\citenamefont {Gabrielse}\ \emph {et~al.}(1990)\citenamefont
  {Gabrielse}, \citenamefont {Fei}, \citenamefont {Orozco}, \citenamefont
  {Tjoelker}, \citenamefont {Haas}, \citenamefont {Kalinowsky}, \citenamefont
  {Trainor},\ and\ \citenamefont {Kells}}]{gabrielse1990thousandfold}%
  \BibitemOpen
  \bibfield  {author} {\bibinfo {author} {\bibfnamefont {G.}~\bibnamefont
  {Gabrielse}}, \bibinfo {author} {\bibfnamefont {X.}~\bibnamefont {Fei}},
  \bibinfo {author} {\bibfnamefont {L.}~\bibnamefont {Orozco}}, \bibinfo
  {author} {\bibfnamefont {R.}~\bibnamefont {Tjoelker}}, \bibinfo {author}
  {\bibfnamefont {J.}~\bibnamefont {Haas}}, \bibinfo {author} {\bibfnamefont
  {H.}~\bibnamefont {Kalinowsky}}, \bibinfo {author} {\bibfnamefont
  {T.}~\bibnamefont {Trainor}}, \ and\ \bibinfo {author} {\bibfnamefont
  {W.}~\bibnamefont {Kells}},\ }\href@noop {} {\bibfield  {journal} {\bibinfo
  {journal} {Physical review letters}\ }\textbf {\bibinfo {volume} {65}},\
  \bibinfo {pages} {1317} (\bibinfo {year} {1990})}\BibitemShut {NoStop}%
\bibitem [{\citenamefont {Li}\ \emph {et~al.}(2011)\citenamefont {Li},
  \citenamefont {Kheifets},\ and\ \citenamefont {Raizen}}]{li2011millikelvin}%
  \BibitemOpen
  \bibfield  {author} {\bibinfo {author} {\bibfnamefont {T.}~\bibnamefont
  {Li}}, \bibinfo {author} {\bibfnamefont {S.}~\bibnamefont {Kheifets}}, \ and\
  \bibinfo {author} {\bibfnamefont {M.}~\bibnamefont {Raizen}},\ }\href@noop {}
  {\bibfield  {journal} {\bibinfo  {journal} {Nat. Phys.}\ }\textbf {\bibinfo
  {volume} {7}},\ \bibinfo {pages} {527} (\bibinfo {year} {2011})}\BibitemShut
  {NoStop}%
\bibitem [{\citenamefont {Gieseler}\ \emph {et~al.}(2012)\citenamefont
  {Gieseler}, \citenamefont {Deutsch}, \citenamefont {Quidant},\ and\
  \citenamefont {Novotny}}]{gieseler2012subkelvin}%
  \BibitemOpen
  \bibfield  {author} {\bibinfo {author} {\bibfnamefont {J.}~\bibnamefont
  {Gieseler}}, \bibinfo {author} {\bibfnamefont {B.}~\bibnamefont {Deutsch}},
  \bibinfo {author} {\bibfnamefont {R.}~\bibnamefont {Quidant}}, \ and\
  \bibinfo {author} {\bibfnamefont {L.}~\bibnamefont {Novotny}},\ }\href@noop
  {} {\bibfield  {journal} {\bibinfo  {journal} {Physical review letters}\
  }\textbf {\bibinfo {volume} {109}},\ \bibinfo {pages} {103603} (\bibinfo
  {year} {2012})}\BibitemShut {NoStop}%
\bibitem [{\citenamefont {Kiesel}\ \emph {et~al.}(2013)\citenamefont {Kiesel},
  \citenamefont {Blaser}, \citenamefont {Deli{\'c}}, \citenamefont {Grass},
  \citenamefont {Kaltenbaek},\ and\ \citenamefont
  {Aspelmeyer}}]{kiesel2013cavity}%
  \BibitemOpen
  \bibfield  {author} {\bibinfo {author} {\bibfnamefont {N.}~\bibnamefont
  {Kiesel}}, \bibinfo {author} {\bibfnamefont {F.}~\bibnamefont {Blaser}},
  \bibinfo {author} {\bibfnamefont {U.}~\bibnamefont {Deli{\'c}}}, \bibinfo
  {author} {\bibfnamefont {D.}~\bibnamefont {Grass}}, \bibinfo {author}
  {\bibfnamefont {R.}~\bibnamefont {Kaltenbaek}}, \ and\ \bibinfo {author}
  {\bibfnamefont {M.}~\bibnamefont {Aspelmeyer}},\ }\href@noop {} {\bibfield
  {journal} {\bibinfo  {journal} {Proceedings of the National Academy of
  Sciences}\ }\textbf {\bibinfo {volume} {110}},\ \bibinfo {pages} {14180}
  (\bibinfo {year} {2013})}\BibitemShut {NoStop}%
\bibitem [{\citenamefont {Asenbaum}\ \emph {et~al.}(2013)\citenamefont
  {Asenbaum}, \citenamefont {Kuhn}, \citenamefont {Nimmrichter}, \citenamefont
  {Sezer},\ and\ \citenamefont {Arndt}}]{asenbaum2013cavity}%
  \BibitemOpen
  \bibfield  {author} {\bibinfo {author} {\bibfnamefont {P.}~\bibnamefont
  {Asenbaum}}, \bibinfo {author} {\bibfnamefont {S.}~\bibnamefont {Kuhn}},
  \bibinfo {author} {\bibfnamefont {S.}~\bibnamefont {Nimmrichter}}, \bibinfo
  {author} {\bibfnamefont {U.}~\bibnamefont {Sezer}}, \ and\ \bibinfo {author}
  {\bibfnamefont {M.}~\bibnamefont {Arndt}},\ }\href@noop {} {\bibfield
  {journal} {\bibinfo  {journal} {Nature communications}\ }\textbf {\bibinfo
  {volume} {4}} (\bibinfo {year} {2013})}\BibitemShut {NoStop}%
\bibitem [{\citenamefont {Millen}\ \emph {et~al.}(2014)\citenamefont {Millen},
  \citenamefont {Fonseca}, \citenamefont {Mavrogordatos}, \citenamefont
  {Monteiro},\ and\ \citenamefont {Barker}}]{millen2014optomechanical}%
  \BibitemOpen
  \bibfield  {author} {\bibinfo {author} {\bibfnamefont {J.}~\bibnamefont
  {Millen}}, \bibinfo {author} {\bibfnamefont {P.}~\bibnamefont {Fonseca}},
  \bibinfo {author} {\bibfnamefont {T.}~\bibnamefont {Mavrogordatos}}, \bibinfo
  {author} {\bibfnamefont {T.}~\bibnamefont {Monteiro}}, \ and\ \bibinfo
  {author} {\bibfnamefont {P.}~\bibnamefont {Barker}},\ }\href@noop {}
  {\bibfield  {journal} {\bibinfo  {journal} {arXiv preprint arXiv:1407.3595}\
  } (\bibinfo {year} {2014})}\BibitemShut {NoStop}%
\bibitem [{\citenamefont {Morigi}\ \emph {et~al.}(2007)\citenamefont {Morigi},
  \citenamefont {Pinkse}, \citenamefont {Kowalewski},\ and\ \citenamefont
  {de~Vivie-Riedle}}]{morigi2007cavity}%
  \BibitemOpen
  \bibfield  {author} {\bibinfo {author} {\bibfnamefont {G.}~\bibnamefont
  {Morigi}}, \bibinfo {author} {\bibfnamefont {P.~W.}\ \bibnamefont {Pinkse}},
  \bibinfo {author} {\bibfnamefont {M.}~\bibnamefont {Kowalewski}}, \ and\
  \bibinfo {author} {\bibfnamefont {R.}~\bibnamefont {de~Vivie-Riedle}},\
  }\href@noop {} {\bibfield  {journal} {\bibinfo  {journal} {Physical review
  letters}\ }\textbf {\bibinfo {volume} {99}},\ \bibinfo {pages} {073001}
  (\bibinfo {year} {2007})}\BibitemShut {NoStop}%
\bibitem [{\citenamefont {Seletskiy}\ \emph {et~al.}(2010)\citenamefont
  {Seletskiy}, \citenamefont {Melgaard}, \citenamefont {Bigotta}, \citenamefont
  {Di~Lieto}, \citenamefont {Tonelli},\ and\ \citenamefont
  {Sheik-Bahae}}]{seletskiy2010laser}%
  \BibitemOpen
  \bibfield  {author} {\bibinfo {author} {\bibfnamefont {D.~V.}\ \bibnamefont
  {Seletskiy}}, \bibinfo {author} {\bibfnamefont {S.~D.}\ \bibnamefont
  {Melgaard}}, \bibinfo {author} {\bibfnamefont {S.}~\bibnamefont {Bigotta}},
  \bibinfo {author} {\bibfnamefont {A.}~\bibnamefont {Di~Lieto}}, \bibinfo
  {author} {\bibfnamefont {M.}~\bibnamefont {Tonelli}}, \ and\ \bibinfo
  {author} {\bibfnamefont {M.}~\bibnamefont {Sheik-Bahae}},\ }\href@noop {}
  {\bibfield  {journal} {\bibinfo  {journal} {Nature Photonics}\ }\textbf
  {\bibinfo {volume} {4}},\ \bibinfo {pages} {161} (\bibinfo {year}
  {2010})}\BibitemShut {NoStop}%
\bibitem [{\citenamefont {Chan}\ \emph {et~al.}(2011)\citenamefont {Chan},
  \citenamefont {Alegre}, \citenamefont {Safavi-Naeini}, \citenamefont {Hill},
  \citenamefont {Krause}, \citenamefont {Groeblacher}, \citenamefont
  {Aspelmeyer},\ and\ \citenamefont {Painter}}]{chan2011laser}%
  \BibitemOpen
  \bibfield  {author} {\bibinfo {author} {\bibfnamefont {J.}~\bibnamefont
  {Chan}}, \bibinfo {author} {\bibfnamefont {T.}~\bibnamefont {Alegre}},
  \bibinfo {author} {\bibfnamefont {A.}~\bibnamefont {Safavi-Naeini}}, \bibinfo
  {author} {\bibfnamefont {J.}~\bibnamefont {Hill}}, \bibinfo {author}
  {\bibfnamefont {A.}~\bibnamefont {Krause}}, \bibinfo {author} {\bibfnamefont
  {S.}~\bibnamefont {Groeblacher}}, \bibinfo {author} {\bibfnamefont
  {M.}~\bibnamefont {Aspelmeyer}}, \ and\ \bibinfo {author} {\bibfnamefont
  {O.}~\bibnamefont {Painter}},\ }\href@noop {} {\bibfield  {journal} {\bibinfo
   {journal} {Nature}\ }\textbf {\bibinfo {volume} {478}},\ \bibinfo {pages}
  {89} (\bibinfo {year} {2011})}\BibitemShut {NoStop}%
\bibitem [{\citenamefont {Aspelmeyer}\ \emph {et~al.}(2013)\citenamefont
  {Aspelmeyer}, \citenamefont {Kippenberg},\ and\ \citenamefont
  {Marquardt}}]{aspelmeyer2013cavity}%
  \BibitemOpen
  \bibfield  {author} {\bibinfo {author} {\bibfnamefont {M.}~\bibnamefont
  {Aspelmeyer}}, \bibinfo {author} {\bibfnamefont {T.~J.}\ \bibnamefont
  {Kippenberg}}, \ and\ \bibinfo {author} {\bibfnamefont {F.}~\bibnamefont
  {Marquardt}},\ }\href@noop {} {\bibfield  {journal} {\bibinfo  {journal}
  {arXiv preprint arXiv:1303.0733}\ } (\bibinfo {year} {2013})}\BibitemShut
  {NoStop}%
\bibitem [{\citenamefont {Ashkin}(2000)}]{ashkin2000history}%
  \BibitemOpen
  \bibfield  {author} {\bibinfo {author} {\bibfnamefont {A.}~\bibnamefont
  {Ashkin}},\ }\href@noop {} {\bibfield  {journal} {\bibinfo  {journal}
  {Selected Topics in Quantum Electronics, IEEE Journal of}\ }\textbf {\bibinfo
  {volume} {6}},\ \bibinfo {pages} {841} (\bibinfo {year} {2000})}\BibitemShut
  {NoStop}%
\bibitem [{\citenamefont {Gieseler}\ \emph {et~al.}(2013)\citenamefont
  {Gieseler}, \citenamefont {Novotny},\ and\ \citenamefont
  {Quidant}}]{gieseler2013thermal}%
  \BibitemOpen
  \bibfield  {author} {\bibinfo {author} {\bibfnamefont {J.}~\bibnamefont
  {Gieseler}}, \bibinfo {author} {\bibfnamefont {L.}~\bibnamefont {Novotny}}, \
  and\ \bibinfo {author} {\bibfnamefont {R.}~\bibnamefont {Quidant}},\
  }\href@noop {} {\bibfield  {journal} {\bibinfo  {journal} {Nature Physics}\ }
  (\bibinfo {year} {2013})}\BibitemShut {NoStop}%
\bibitem [{\citenamefont {Bateman}\ \emph {et~al.}(2013)\citenamefont
  {Bateman}, \citenamefont {Nimmrichter}, \citenamefont {Hornberger},\ and\
  \citenamefont {Ulbricht}}]{bateman2013near}%
  \BibitemOpen
  \bibfield  {author} {\bibinfo {author} {\bibfnamefont {J.}~\bibnamefont
  {Bateman}}, \bibinfo {author} {\bibfnamefont {S.}~\bibnamefont
  {Nimmrichter}}, \bibinfo {author} {\bibfnamefont {K.}~\bibnamefont
  {Hornberger}}, \ and\ \bibinfo {author} {\bibfnamefont {H.}~\bibnamefont
  {Ulbricht}},\ }\href@noop {} {\bibfield  {journal} {\bibinfo  {journal}
  {Nature Communications}\ }\textbf {\bibinfo {volume} {5}},\ \bibinfo {pages}
  {4588} (\bibinfo {year} {2013})}\BibitemShut {NoStop}%
\bibitem [{\citenamefont {Romero-Isart}\ \emph {et~al.}(2011)\citenamefont
  {Romero-Isart}, \citenamefont {Clemente}, \citenamefont {Navau},
  \citenamefont {Sanchez},\ and\ \citenamefont {Cirac}}]{Romero2012magneto}%
  \BibitemOpen
  \bibfield  {author} {\bibinfo {author} {\bibfnamefont {O.}~\bibnamefont
  {Romero-Isart}}, \bibinfo {author} {\bibfnamefont {L.}~\bibnamefont
  {Clemente}}, \bibinfo {author} {\bibfnamefont {C.}~\bibnamefont {Navau}},
  \bibinfo {author} {\bibfnamefont {A.}~\bibnamefont {Sanchez}}, \ and\
  \bibinfo {author} {\bibfnamefont {J.~I.}\ \bibnamefont {Cirac}},\ }\href@noop
  {} {\bibfield  {journal} {\bibinfo  {journal} {ArXiv e-prints}\ } (\bibinfo
  {year} {2011})},\ \Eprint {http://arxiv.org/abs/1112.5609} {1112.5609
  [quant-ph]} \BibitemShut {NoStop}%
\bibitem [{\citenamefont {Coccia}\ \emph {et~al.}(1998)\citenamefont {Coccia},
  \citenamefont {Fafone}, \citenamefont {Lobo},\ and\ \citenamefont
  {Ortega}}]{Coccia:1998}%
  \BibitemOpen
  \bibfield  {author} {\bibinfo {author} {\bibfnamefont {E.}~\bibnamefont
  {Coccia}}, \bibinfo {author} {\bibfnamefont {V.}~\bibnamefont {Fafone}},
  \bibinfo {author} {\bibfnamefont {J.~A.}\ \bibnamefont {Lobo}}, \ and\
  \bibinfo {author} {\bibfnamefont {J.~A.}\ \bibnamefont {Ortega}},\
  }\href@noop {} {\bibfield  {journal} {\bibinfo  {journal} {Phys. Rev. D}\
  }\textbf {\bibinfo {volume} {57}},\ \bibinfo {pages} {2051} (\bibinfo {year}
  {1998})}\BibitemShut {NoStop}%
\bibitem [{\citenamefont {Brandt}(1998)}]{Brandt:1989}%
  \BibitemOpen
  \bibfield  {author} {\bibinfo {author} {\bibfnamefont {E.~H.}\ \bibnamefont
  {Brandt}},\ }\href@noop {} {\bibfield  {journal} {\bibinfo  {journal}
  {Science}\ }\textbf {\bibinfo {volume} {243}},\ \bibinfo {pages} {349}
  (\bibinfo {year} {1998})}\BibitemShut {NoStop}%
\end{thebibliography}%

\end{document}